\documentclass[graybox]{svmult}

\usepackage{mathptmx}       
\usepackage{helvet}         
\usepackage{courier}        
\usepackage{type1cm}        
\usepackage{makeidx}         
\usepackage{graphicx}        
\usepackage{multicol}        
\usepackage[bottom]{footmisc}

\usepackage{color}
\usepackage{amsmath}
\usepackage[space]{grffile}
\usepackage{textcomp}
\definecolor{keywordcolor}{rgb}{0,0,1}

\makeindex             

\begin{document}

\title*{Quantum Light Storage in Solid State Atomic Ensembles.}

\author{ Hugues de
Riedmatten, Mikael Afzelius}

\institute{Hugues de Riedmatten \at ICFO-The Institute of Photonic
Sciences, Barcelona, Spain \at ICREA-Instituci\'{o} Catalana de Recerca i Estudis Avan\c cats, 08015 Barcelona, Spain\\ \email{hugues.deriedmatten@icfo.es}
\and Mikael Afzelius \at Group of Applied Physics, University of
Geneva, Switzerland \email{mikael.afzelius@unige.ch}}

\maketitle

\abstract{In this chapter, we will describe the storage and
retrieval of quantum light (heralded single photons and entangled
photons) in atomic ensembles in a solid state environment.  We
will consider ensembles of rare-earth ions embedded in dielectric
crystals.  We will describe the methods used to create quantum
light spectrally compatible with the narrow atomic transitions, as
well as possible protocols based on dipole rephasing that can be
used to reversibly map the quantum light onto collective atomic
excitations.  We will review the experimental state of the art and
describe in more detail quantum light storage experiments in
neodymium and praseodymium doped crystals.}

\section{Introduction}
\label{Sec:Intro}

Harnessing strong and coherent interactions between quantum light
and matter is an important ability in quantum science. These
interactions can be used to build light-matter interfaces enabling
reversible quantum state transfer between photons and atoms. One
important application of these interfaces is the realization of
photonic quantum memories
\cite{Hammerer2010,Lvovsky2009,Simon2010,Bussieres2013} which
allow storage of quantum information carried by photons. Quantum
memories are important devices in quantum information science
because they can be used as synchronization devices when many
different probabilistic quantum processes are linked together.
They are therefore required for scalable protocols using photons,
with potential applications in optical quantum computing,
generation of multiphoton states from probabilistic pair sources
\cite{Nunn2013}, quantum information networks \cite{Kimble2008}
and long-distance quantum communication using quantum repeaters
\cite{Briegel1998,Duan2001,Sangouard2011}.

The realization of quantum memories for light requires efficient
and reversible mapping of photons onto long lived atomic
coherences. This in turn requires strong interactions between
light and matter. However, in free space the interaction between a
single photon and a single atom is usually weak. One way to
overcome the problem is to place the atom in a high finesse cavity
which strongly enhances the interaction \cite{Specht2011} (see
chapter (Kuhn)). Another way is to use a collection of atoms,
where the atom light coupling is enhanced by a factor of
$\sqrt{N}$ with $N$ being the number of atoms involved. Single
photons are stored in atomic ensembles as collective atomic
excitation, sometimes called superatoms. These superatoms have the
important property that they can be efficiently converted to
single photons in a well defined spatio-temporal mode thanks to a
collective interference between all the involved emitters
\cite{Duan2001}. This so-called collective enhancement is at the
heart of most quantum memory protocols in atomic ensembles. The
quantum control of collective atomic coherences is therefore a key
task in the field of quantum memories.

For applications involving transfer of quantum information over
large distances, remote quantum memories must be entangled
\cite{Duan2001,Simon2007a}. This requires that the remote quantum
memories must exchange quantum information using e.g. single
photons, or that photons emitted by the quantum memories interfere
at a central location between the two quantum memories. If optical
fibers are used, this means that quantum memories must be
connected to the optical fiber network, in particular to photons
at telecom wavelengths in order to minimize optical losses in the
fiber transmission.

Quantum memories for light were first demonstrated in atomic
gases, both room-temperature gases and ensembles of laser-cooled
atoms. Several review papers can also be found on that subject
\cite{Hammerer2010,Sangouard2011}. Some solid-state systems offer
interesting perspectives as quantum memories for light, such as
rare-earth doped crystals
\cite{Riedmatten2008,Hedges2010,Clausen2011,Saglamyurek2011,Rielander2014},
nitrogen-vacancy centers in diamond \cite{Togan2010,Bernien2013},
phonons in diamond \cite{Lee2012,England2013} and quantum dots
\cite{deGreve2012,Bao2012,Simon2010} (see also chapter McMahon). Realizing quantum memories
in solid-state systems would, in general, have several advantages,
such as the absence of atomic motion and the prospects for
integrated devices, which may facilitate large-scale deployment of
these techniques in future quantum networks. But, controlling
light-matter interactions in solid-state materials also poses
important challenges, such as preserving the quantum coherence in
a solid-state environment.

Here we will discuss rare-earth-ion doped crystals for quantum
memories. These crystals provide a large number of atoms naturally
trapped in a solid-state matrix, with spectrally narrow optical
and spin transitions. Due to their particular electron level
structure, they also provide exceptional coherence properties,
both for the optical and spin transitions, when cooled to
cryogenic temperatures.

This chapter will describe experiments and techniques developed to
store non-classical light in rare-earth-ion doped crystals. In
section \ref{Sec:RECrystals}, we describe in more details the
relevant properties of rare-earth-ion doped crystals and the
reasons why these are interesting materials for quantum light
storage. In section \ref{Sec:Protocols}, we describe quantum
memory protocols that have been proposed to store quantum
information in doped crystals. In section
\ref{Sec:State_of_the_art}, we review the experimental
state-of-the-art of storing non-classical states of light in
crystals. In the following two sections we describe the
development of two specific sources of non-classical light
(Section \ref{Sec:SPDCsource:General}) and their application to
non-classical light storage (Section \ref{Sec:QStorage}) in
Pr$^{3+}$ and Nd$^{3+}$ ions doped crystals. Finally, in section
\ref{Sec:prospects}, we comment on the prospects for extending
quantum light storage to longer storage times  in these systems.

\section{Rare-earth-ion doped crystals}

 \label{Sec:RECrystals}
 The
energy structure of ions in solid-state materials is usually
strongly affected by the lattice of the host crystal, resulting in
broad optical transitions with very short optical coherence time.
Striking exceptions to this are rare-earth ion impurities in
crystals, whose 4f-4f transitions were found to be extremely
narrow when the first high-resolution resonance spectra were
obtained in the 1970s \cite{Macfarlane2002}. The sharp lines are
due to the shielding of the 4f electron shell from outermost 5s
and 5p electrons, which reduces the coupling of the 4f electrons
to the lattice. This explains the atomic-like properties of the
lanthanides in a crystal. There exist several excellent books
\cite{Macfarlane1987,Liu2005} and reviews of optical properties of
rare-earth-ion doped crystals
\cite{Macfarlane2002,Macfarlane1990,Macfarlane2007} and their
application in quantum information science
\cite{Tittel2010,Thiel2011}. Here we will summarize some
properties that are particularly relevant for quantum memory
applications.

The effective shielding of the 4f electron shell results in
extremely narrow homogeneous and inhomogeneous line widths of the
radiative 4f-4f transitions in rare-earth doped crystals. At
cryogenic temperatures, the inhomogeneous broadening is the
dominating broadening process, analogous to the Doppler broadening
of room temperature alkali gases. But in contrast to the dynamical
Doppler broadening, where atoms jump between velocity classes due
to velocity changing collisions, the inhomogeneous broadening of
rare-earth ion doped crystals is to a large degree static. This is
a result of the physical origin of the inhomogeneous broadening
\cite{STONEHAM1969}, which can be due to local crystal strain or
interactions between dopants \cite{Sellars2004}. In some cases one
can observe a time-dependent broadening of a spectral channel over
time, which is known as spectral diffusion \cite{Bottger2006a}.
However, this effect is usually weak and is taken into account by
a time-dependent homogeneous line width \cite{Bottger2006a}. The
static inhomogeneous broadening and the large number of spectral
channels that can be manipulated with precise lasers have
important consequences for quantum storage experiments. These
features are used in the quantum memory schemes specifically
developed for these materials, resulting in capabilities difficult
to obtain in gas phase experiments.

The optical inhomogeneous linewidth varies strongly between
crystal hosts. In some cases the degree of broadening can be
related to the ionic radius mismatch between the rare-earth dopant
ion and the lattice ion that it replaces. Usual inhomogeneous
broadenings range from a few hundreds of MHz to tens of GHz,
although extreme values of around 10 MHz \cite{Macfarlane1998} and
250 GHz have been observed \cite{Sun2002}.

The optical homogeneous linewidth can be extremely narrow (ranging
from $<$1 kHz to 1 MHz) if the sample is cooled to temperatures
below 5-10 K \cite{Macfarlane1987,Liu2005}. Above this
approximative temperature range, the homogeneous linewidth usually
increases rapidly, displaying a $T^7$ or $T^9$ dependence, due to
coupling to phonons (spin-lattice relaxation). Below this
temperature range, due to the shielding by the 5s and 5p
electrons, the homogeneous linewidth is often limited by magnetic
interactions with other rare-earth ion dopants or magnetic
constituents of the lattice. This has led to the general
understanding that crystal materials with low nuclear spin
concentration \cite{Macfarlane1981,Yano1991} and low rare-earth
ion dopant concentration \cite{Bottger2006a} provides a way of
obtaining long coherence times, both for optical and hyperfine
transitions. To increase the coherence times of hyperfine levels,
it was also realized that one can exploit the non-linear magnetic
Zeeman effect of the hyperfine levels in order to find sweet spots
where there is a zero first-order Zeeman effect (ZEFOZ) of the
hyperfine transition \cite{Fraval2004}. This effectively decouples
the hyperfine transition from the fluctuating magnetic
environment, which can increase the hyperfine coherence time with
orders of magnitude, akin to clock transitions used in alkali
atoms. In addition, one can apply dynamical decoupling schemes to
further increase the spin coherence time
\cite{Fraval2005,Pascual-Winter2012,Lovric2013,Heinze2013}. A recent experiment demonstrated a coherence time of 6 hours in 
Eu$^{3+}$:Y$_2$SiO$_5$  \cite{Zhong2015}.

The parity-forbidden 4f-4f transitions are only weakly allowed in
crystals, and arise due to admixtures of excited configurations of
different parity to the 4f configuration. As a consequence
radiative lifetimes are long, usually in the range of 100 $\mu$s
to 10 ms. The oscillator strengths of transitions relevant to
quantum information applications are in the range of 10$^{-8}$ to
10$^{-6}$ \cite{McAuslan2009}. The weak absorption probability of
individual ions, as compared to the alkali D$_1$ and D$_2$ lines,
is compensated by the high density of ions typical for a crystal.
The doping concentrations are in the range of 10 to 1000 pm,
resulting in typical number densities in the range of 10$^{17}$ to
10$^{19}$ ions/cm$^{3}$. The absorption coefficients for typical
doping levels range from $\alpha$ = 1 to 50 cm$^{-1}$, which often
make the transitions close to opaque for a 1 cm long crystal.

We finally comment on the electronic ground-state substructure,
which is relevant for quantum information processing in general.
The RE$^{3+}$ ions can be divided into Kramers and non-Kramers
ions, which have odd or even number of electrons, respectively.
This is relevant since the two groups react differently to the
interaction with the surrounding lattice ions (crystal-field
interaction). In a low-symmetry RE$^{3+}$ doping site, the crystal
field interaction completely lifts the electronic ground-state
degeneracy for non-Kramers ions. For Kramers ions the
crystal-field interaction results gives rise to a series of
degenerate doublets, which is due to Kramers time-reversal
symmetry. As a consequence the Kramers ions have a strong magnetic
dipole moment (order of the Bohr magneton), while non-Kramers ions
usually have weak nuclear magnetic moments due to the quenching of
the electronic magnetic moment. In some cases, however, the
nuclear moment can be strongly enhanced in non-Kramers ions
through interactions induced by the crystal-field Hamiltonian. In
addition we also need to consider different types of hyperfine
interactions. As an example, in Eu$^{3+}$ and Pr$^{3+}$ doped
crystals, which are non-Kramers ions, these ions have three
hyperfine states, due to quadrupole-type interactions
\cite{Macfarlane1987,Liu2005}. The hyperfine level separations in
these ions are of the order of 10 to 100 MHz. In Kramers ions,
such as Er$^{3+}$ and Nd$^{3+}$, a spectrally resolved Zeeman
ground-state doublet can be formed by applying a rather weak field
($<$1 Tesla) \cite{Sun2008}. For isotopes with non-zero nuclear
spin, the interaction with unquenched electron spin results in a
strong hyperfine interaction, of the order of 1 GHz for Kramers
ions \cite{Guillot-Noel2006}. These different considerations are
important for different aspects of quantum memory applications,
since they affect the frequencies used for state preparation of
the ions and possible limitations of the memory bandwidth due to
interfering optical-hyperfine transitions.
\section{Quantum memory protocols}
\label{Sec:Protocols}

In this section we will shortly discuss quantum memory protocols
that can be used in rare-earth-ion doped crystals. We will
particularly discuss the atomic frequency comb protocol, which so
far is the only protocol that has been used in these systems for
storage of quantum states of light. For the reader who wants an
overview of different quantum memory schemes we refer to the many
excellent reviews that have been published
\cite{Hammerer2010,Lvovsky2009,Tittel2010,Simon2010,Bussieres2013}.

An important class of quantum memory schemes have been inspired by
the photon echo process \cite{Tittel2010}, which is the optical
analogue of the spin echo. In the conventional photon echo process
the optical pulse to be stored is absorbed by an inhomogeneous
ensemble of atoms, typically rare-earth impurity ions in a
crystal. The induced atomic coherence then undergo inhomogeneous
dephasing, but the dephasing can be reversed by applying an
optical $\pi$-pulse a time $\tau$ after the input pulse. After a
time 2$\tau$ the atomic coherences are back in phase, which
results in a strong collective emission known as a photon echo.
Photon echo processes have been investigated for storing and
processing coherent states of light since the 1980s
\cite{Tittel2010}. It was therefore natural to consider if the
same processes and materials could be used to store the quantum
state of a single photon. It was realized, however, that the
optical $\pi$-pulse used in conventional photon echoes would cause
too much spontaneous emission noise, due to the high degree of
atomic excitation induced by it \cite{Ruggiero2009}. In 2001
Moiseev and Kr\"oll \cite{Moiseev2001} made an initial proposal
for a noise free photon-echo quantum memory scheme. The scheme was
proposed for Doppler-broadened lines, but sparked the interest for
finding a similar scheme adapted to inhomogeneously broadened
solid-state ensembles. Around 2005-2006 Nilsson and Kr\"{o}ll
\cite{Nilsson2005}, Kraus et al. \cite{Kraus2006} and Alexander et
al. \cite{Alexander2006} proposed a quantum memory scheme where
the inhomogeneous dephasing was controlled not by a strong
$\pi$-pulse, but by an external electric field gradient. The basic
idea is to create a narrow spectral feature using optical pumping
techniques, which is then broadened using an external field. The
scheme was coined controlled reversible inhomogeneous broadening
(CRIB), while a later modified version of the scheme was named
gradient echo memory (GEM) \cite{Hetet2008}. Several experimental
realizations of the CRIB/GEM storage scheme in rare-earth-ion
doped crystals followed \cite{Alexander2006}, \cite{Hetet2008a},
including storage of weak coherent states at the single photon
level \cite{Hedges2010}, \cite{Lauritzen2010}. It is worth noting
that the GEM experiment reported by Hedges et al.
\cite{Hedges2010} demonstrated one of the highest efficiencies reached
in any quantum memory, 69\%, and the highest reached in a
solid-state memory. In rare-earth-ion doped crystals the CRIB/GEM
experiments were implemented using external electric fields, based
on the linear Stark shift. In some systems, however, the linear
Stark shift is zero due to symmetry considerations
\cite{Chaneliere2008}.

\begin{figure}[b]

    \includegraphics[scale=.60]{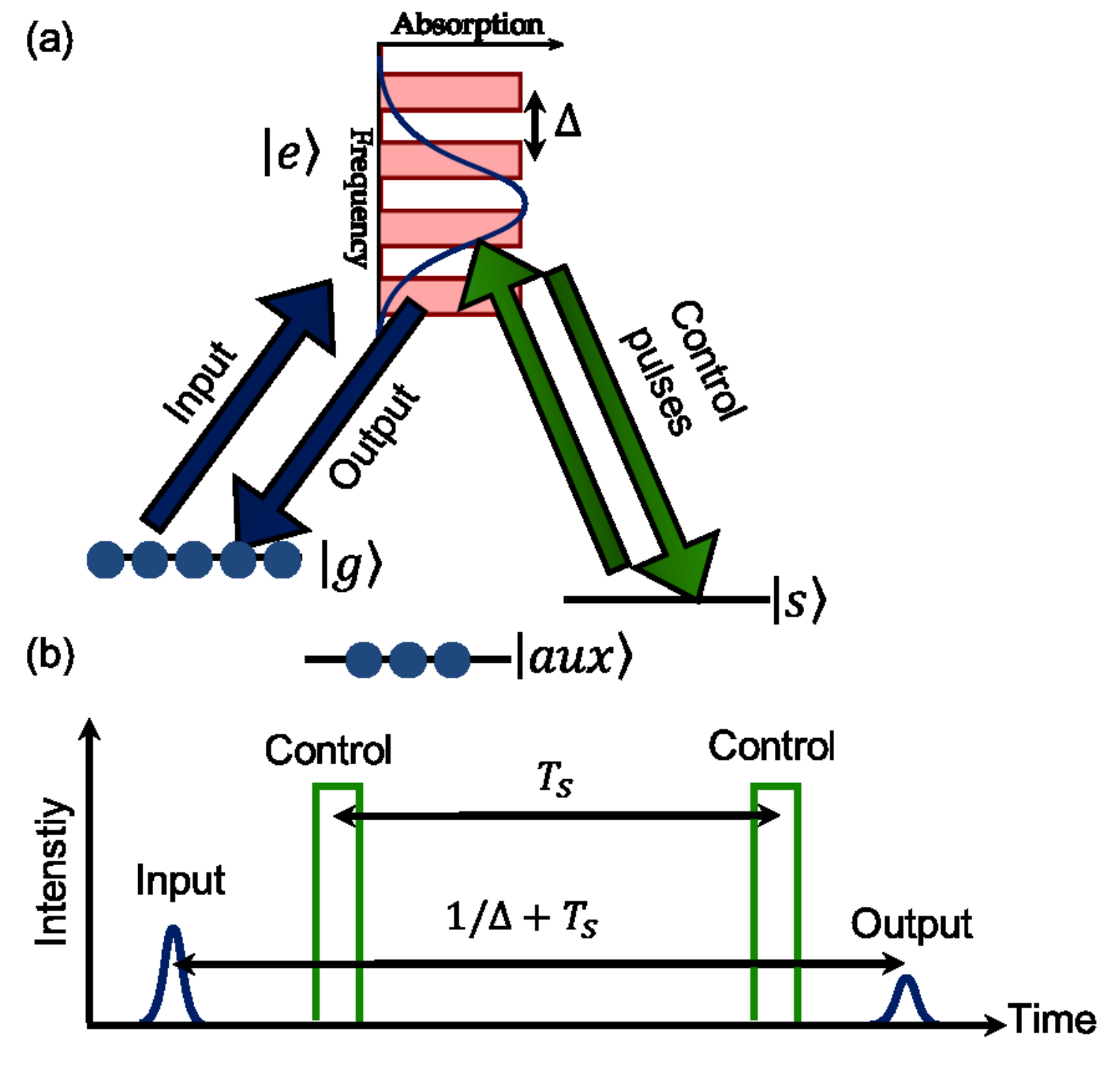}

    \caption{(a) The input pulse is absorbed on a strongly absorbing
transition whose inhomogeneous spectrum is shaped into a comb with
periodicity $\Delta$. After the absorption of the input, a control
pulse converts the initial optical coherence into a spin
coherence, see panel (b) for the timing. Another control pulse
applied a time $T_S$ after the first one re-establishes the
optical coherence, which evolves towards an echo emission after a
total storage time $1/\Delta+T_S$. This scheme is the complete AFC
spin-wave memory. If the control pulses are not applied, then the
input will give rise to an output echo after a total storage time
$1/\Delta$, called the AFC echo scheme. We refer to the text for
more details on the processes and the required energy structure.}

    \label{fig:AFCscheme}

\end{figure}

In 2008 the atomic frequency comb (AFC) quantum memory scheme was
proposed \cite{Riedmatten2008,Afzelius2009}. The motivation for
the scheme was the storage of trains of pulses, so-called temporal
multimode storage, which turned out to be difficult using the CRIB
scheme due to the scaling of the number of modes as a function of
the optical depth of the storage material. In CRIB memories the
number of modes scales linearly with the optical depth $d$ of the
transition \cite{Simon2007a}, while for AFC it is independent of
optical depth, although depending on other critical parameters, as
will be discussed later. For electro-magnetically induced
transparency (EIT) and Raman techniques, two common light-storage
techniques, the number of modes scales as $\sqrt{d}$
\cite{Nunn2008a}.

The AFC scheme is based on a spectral tailoring of the
inhomogeneous absorption spectrum of an optical transition
$|g\rangle - |e\rangle$, where one ideally wants to make a
periodic series of narrow, highly absorbing peaks. This forms the
atomic frequency comb, which can be characterized by its spacing
$\Delta$ and finesse $F$, in analogy with an optical cavity. A
single-photon state, with a bandwidth $\gamma_p$ larger than
$\Delta$, can be completely absorbed by the comb. Indeed, if the
input pulse is short enough (meaning $\gamma_p > \Delta$), the
effect of the short interaction in time is a spectral averaging of
the sharp AFC structure into a smooth distribution, due to Fourier
arguments, allowing for uniform absorption over the photons
bandwidth. The effective optical depth of the comb is roughly
$\tilde{d}=d/F$ \cite{Afzelius2009}, depending on the shape of the
peaks, where $d$ is the peak absorption depth. The absorption
probability is $1-\exp(-\tilde{d})$, showing that complete
absorption can be achieved for high enough $d$ for any finesse
$F$. For a more complete description, we refer the reader to
\cite{Afzelius2009}.

Conditioned on the absorption of a single photon, the atomic state
can be described by a collective Dicke state $\sum_k c_k
|g\cdot\cdot\cdot e_k \cdot\cdot\cdot g\rangle$ \cite{Dicke1954},
where the amplitudes $c_k$ depend on the detuning and spatial
position of the particular atom $k$. These modes are initially in
phase, but the collective state will rapidly dephase into a
non-collective state $\sum_k \exp(-i2\pi\delta_kt) c_k
|g\cdot\cdot\cdot e_k \cdot\cdot\cdot g \rangle$, since each term
acquires an individual phase depending on the detuning $\delta_k$
of each excited atom. If we consider an AFC having very sharp
peaks, then the detunings $\delta_k$ are approximately a discrete
set such that $\delta_k = m_k\Delta$, where $m_k$ are integers. It
follows that the collective state is re-established after a time
$1/\Delta$, which leads to a coherent photon-echo
\cite{Mossberg1979,Carlson1984,Mitsunaga1991} type re-emission in
the forward spatial mode defined by the absorbed photon. We note that the AFC echo can also be interpreted as a slow-light effect induced by the comb structure \cite{Bonarota2012}.

The scheme described here can only be used as a delay-line, with a
fixed storage time $1/\Delta$. In the following we will refer to
this scheme as an \textit{AFC echo scheme}. A recent quantum
repeater protocol is entirely based, however, on this scheme, but
which requires heavy frequency and time multiplexing to be
efficient \cite{Sinclair2014}. It should also be emphasized that
the AFC echo scheme provides a dynamical delay-line, which can be
re-programmed with a rate related to the comb creation time. The
temporal multimode capacity of the AFC scheme does not depend on
optical depth, as mentioned above. The number of modes that one
can store depends simply on the number of peaks in the comb, which
in turn depends on the ratio of the comb periodicity $\Delta$ to
the total comb bandwidth $\Gamma$.

To be able to read out a AFC memory on demand, the original
proposal was based on a conversion of the optical excitation into
a spin excitation \cite{Afzelius2009}. This can be done by
applying an optical control pulse that transfers the single
optical excitation to a spin state, for instance a $\pi$-pulse,
after the absorption of the single photon, but before the
appearance of the AFC echo. This requires an additional ground
state level $|s\rangle$, such that the states
$|g\rangle$,$|e\rangle$ and $|s\rangle$ form a so-called
$\Lambda$-system. To read out the memory a second control pulse is
applied after a spin-wave storage time $T_S$, after which the
collective Dicke state continues to evolve towards the AFC echo
emission, after a total storage time $T_S+1/\Delta$. In addition
to providing on-demand read out of the memory, it can also provide
a longer total storage time, since the spin coherence time can be
orders of magnitude longer than the optical coherence time. In the
following we will refer to this scheme as a \textit{AFC spin-wave
memory}. The spin-wave storage requires a coherent spin-transition
$|g\rangle$-$|s\rangle$ for storing the spin coherence, but also
an additional state $|aux\rangle$ with long population lifetime.
The auxiliary ground state $|aux\rangle$ is used for storing
population that has been pumped away optically during the AFC
creation process. The state $|aux\rangle$ is also needed in the
AFC echo scheme described above. The need for three ground-state
levels and a coherent spin transition $|g\rangle$-$|s\rangle$
limits the number of known materials that can be used for
spin-wave storage.

The efficiency of the complete AFC spin-wave memory depends on
several factors. The most important one is the efficiency of the
AFC echo, which in turn depends on the optical depth of the
material, the comb parameters and the direction of recall. But one
also needs to consider the efficiency of the optical control
pulses and spin dephasing during the spin-wave storage time $T_S$.
In most cases these factors act independently on the total
efficiency, leading to the simple efficiency formula

\begin{equation}
\eta=\eta_{AFC} \eta_{C}^2 \eta_{S}
\end{equation}

\noindent where $\eta_{AFC}$ is the AFC echo efficiency,
$\eta_{C}$ the efficiency of one optical control pulse and
$\eta_{S}$ accounts for loss of efficiency due to spin
decoherence. We here assume that the control pulses introduce no
decoherence, we only take into account a limited transfer
efficiency of population.

The AFC echo efficiency depends on the direction of recall. In
forward direction the re-absorption effect in a optically dense
medium limits the efficiency to 54\% \cite{Moiseev2004,Sangouard2007}. In
backward recall an interference effect makes it possible to reach
100\% in principle \cite{Sangouard2007,Afzelius2009}. Backward
recall can be achieved by using counter-propagating control
pulses, but then only in spin-wave storage. The AFC echo
efficiency formulas for both cases are given below

\begin{equation}
\eta^{fw}_{AFC}=\tilde{d}^2 \exp(-\tilde{d}) \eta_{deph},
\end{equation}

\begin{equation}
\eta^{bw}_{AFC}=(1-\exp(-\tilde{d}))^2 \eta_{deph},
\end{equation}

\noindent where $\eta_{deph}$ is a dephasing factor that accounts
for the finite width and shape of the AFC teeth. It should be
emphasized that these formulas also apply to the CRIB/GEM scheme.
As shown in Refs \cite{Sangouard2007,Afzelius2009}, $\eta_{deph}$
is simply the Fourier transform of a \textit{single tooth}
function in the comb, evaluated at the time of the AFC echo
$1/\Delta$. The effective absorption depth $\tilde{d}$ also
depends on the exact shape of the AFC teeth. In Ref.
\cite{Afzelius2009} Gaussian peaks where considered, while in Ref.
\cite{Chaneliere2010} formulas were given for Lorentzian shaped
teeth. Later Bonarota \textit{et al.} \cite{Bonarota2010} showed
that square peaks give the highest efficiency for a given peak
optical depth $d$. For square-shaped peaks $\tilde{d}=d/F$ exactly
and $\eta_{deph}=\rm{sinc}^2(\pi/F)$.

The optical depth is in practice the most crucial parameter, which
led to the proposal to
 put the memory in an optical cavity to enhance the effective interaction
length \cite{Moiseev2010a},\cite{Afzelius2010a}. It was shown that
the cavity could be operated in an optimal regime, where the input
mirror reflectivity $R$ of an asymmetric cavity is tuned to the
effective optical depth of the memory $\tilde{d}$ such that
$R=\exp(-2 \tilde{d})$, assuming $\tilde{d}<1$, which is called an
\textit{impedance-matched} cavity. At the impedance-match point
complete absorption can in principle be achieved, if all other
losses are much smaller than  $\tilde{d}$, and the efficiency is
then bounded only by the intrinsic dephasing $\eta_{deph}$. As a
consequence the cavity approach can lead to close to 100\%
efficiency, without resorting to the phase-matching operation
required for backward recall. Note also that the impedance-matched
cavity scheme can be applied to any memory scheme based on control
of the inhomogeneous dephasing, e.g. AFC, CRIB or GEM. Recent
experimental demonstrations of the cavity scheme reached AFC echo
efficiencies with bright pulses of 56\% \cite{Sabooni2013} in
Pr$^{3+}$:Y$_2$SiO$_5$  and 53\% \cite{Jobez2014} in Eu$^{3+}$:Y$_2$SiO$_5$, the highest reported
AFC efficiencies to date.

\section{State of the art}

\label{Sec:State_of_the_art}

Although the focus of this chapter is the interaction of quantum
light with rare-earth doped solids, we first review a series of
experiments that have been performed with weak coherent states at
the single photon level. This type of experiment allows the
testing of several aspects relevant for quantum light storage, in
particular the coherence preservation, the noise added in the
storage and retrieval processes as well as the waveform
preservation. In addition, although the light at the input is
classical, it has been shown that it is possible to infer the
quantum character of the storage under certain conditions
\cite{Hedges2010,Specht2011,Gundogan2012,Sinclair2014}.

The first demonstration of storage and retrieval of light at the
single photon level in a solid state device, which was also the
first demonstration of the AFC echo scheme, was done in 2008 at
the University of Geneva \cite{Riedmatten2008}. Weak coherent
light pulses were stored for up to 1 $\mu s$ using the atomic
frequency comb scheme in a Nd$^{3+}$:YVO$_4$ crystal. Single
photon level time-bin qubits were also stored and the coherence
was shown to be preserved to a high degree during the storage and
retrieval process. Finally, a proof of principle experiment of
temporal multimodality of the protocol was done, with the storage
and retrieval of 4 temporal modes. The storage and retrieval
efficiency was $< 1 \%$ in that initial demonstration. However,
several other single photon level experiments  in other materials
have since then demonstrated the AFC echo scheme at the single
photon level with much higher efficiencies, reaching 9 $\%$
efficiency in Tm$^{3+}$:YAG \cite{Chaneliere2010} and 25 $\%$ in Pr$^{3+}$:Y$_2$SiO$_5$
\cite{Sabooni2010,Maring2014}.

 Another
aspect that has been improved in recent experiments is the
multimode capacity. The reversible mapping of up to 64 weak pulse
temporal modes has been demonstrated in a Nd:YSO crystal
\cite{Usmani2010}. The coherence was verified by simultaneously
storing and analyzing multiple time-bin qubits. It has also been
shown that, combined with phase modulators, AFC can be used as a
programmable processor for spectral and temporal manipulation of
single qubits \cite{Saglamyurek2014}. Hong-Ou-Mandel interference
between two AFC echoes recalled from Tm doped waveguides has been
demonstrated \cite{Jin2013}. Finally, the AFC storage was also
extended to polarization qubits
\cite{Gundogan2012,Clausen2012,Zhou2012}, and more recently to
spectrally multiplexed time-bin qubits  with selective readout in
frequency \cite{Sinclair2014}. Note that for all AFC echo
experiments mentioned above, the storage was done in the excited
state only, leading to short and pre-determined storage time. The
CRIB/GEM protocol allows on demand read-out even with storage in
the excited state, as has been demonstrated at the single photon
level in an Erbium doped crystal \cite{Lauritzen2010} and in a Pr
doped crystal \cite{Hedges2010}. The latter experiment reported
the highest efficiency for any solid state memory so far (69
$\%$). It also showed that the memory operated in the quantum
regime, meaning that the storage and retrieval fidelity was
measured to be higher than the one achievable with a classical
memory.

The first demonstration of quantum light storage in solid state
device was reported simultaneously in 2011 by two groups, one from
the University of Geneva \cite{Clausen2011} and one from the
University of Calgary \cite{Saglamyurek2011}. Both experiments
demonstrated the storage of entangled photons in a rare-earth
doped crystal, using the AFC echo scheme. Both experiments created
compatible photon pair sources with one photon matching the
storage device and the other photon at telecommunication
wavelength.  The two teams used different storage media with
different properties and bandwidth.  The Calgary experiment used a
broadband AFC in a Tm doped Lithium Niobate waveguide absorbing
light at 793 nm. One photon of the pair was stored for 7 ns in the
waveguide, with a storage and retrieval efficiency of 2 $\%$ (
excluding coupling losses in the waveguide). It was also shown
that time-bin entanglement was preserved during the storage and
retrieval, and a violation of a Bell inequality was demonstrated
between the telecom photon and the stored and retrieved photon.
The Geneva experiment used a Nd doped crystal absorbing at 883 nm,
with a storage bandwidth of 120 MHz and featured a maximal storage
time of 200 ns and a maximal efficiency of 20 $\%$. The
preservation of energy-time entanglement was demonstrated. These
experiments will be described in more detail in section
\ref{Sec:QStorage:Nd}.

These experiments demonstrated for the first time entanglement
between a telecom photon and a collective optical atomic
excitation in a solid state device. It should be noted however
that in both cases the photons were stored as
optical atomic excitations, leading to short and only
pre-determined storage times.

Following these initial experiments, further developments by the
same groups included the storage and retrieval of polarization
\cite{Clausen2012} and time-bin qubits \cite{Saglamyurek2012}
carried by heralded single photons. The Geneva group also reported
an experiment demonstrating entanglement between two crystals
\cite{Usmani2012} (it will be described in more detail in section
\ref{Sec:QStorage:Nd}), followed by a experiment demonstrating
quantum teleportation of a telecom wavelength photon onto a
collective atomic optical excitation \cite{Bussieres2014}. Furthermore, the quantum storage of a 3-dimensional orbital-angular-momentum entangled photon has been reported in a  Nd$^{3+}$:YVO$_4$ crystal by a group in Hefei \cite{Zhou2014}.

In 2014 an experiment demonstrating quantum storage of heralded
single photons using the AFC echo scheme in a Praseodymium doped
crystal absorbing at 606 nm was carried out at ICFO. This material
has demonstrated promising properties in the storage of classical
light, including long storage times up to 1 minute
\cite{Longdell2005,Heinze2013} and high storage and retrieval
efficiencies as mentioned above \cite{Hedges2010}. Contrary to
materials used in previous demonstrations, it also possesses 3
ground state levels, such that spin-wave storage is in principle
possible. However this comes with the drawback that the spacing
between the hyperfine states is small, therefore limiting the
storage bandwidth to a few MHz. This poses strong challenges for
the realization of a suitable quantum light source. The
realization of such a source \cite{Fekete2013} and the storage
experiment \cite{Rielander2014} will be described in more detail
in sections \ref{Sec:SPDCsource:Pr} and \ref{Sec:QStorage:Pr},
respectively.

While the previous experiments have been performed in rare-earth
doped crystals, a recent experiment demonstrated that it is also
possible to store non classical light states in an amorphous
environment. An erbium-doped standard telecom glass fiber was used
as storage device, and photons at telecom wavelengths were stored
using the atomic frequency comb scheme, with an efficiency around
1 $\%$ and a storage time of 5 ns \cite{Saglamyurek2015}. The
experiment also showed that entanglement was preserved during the
storage in the fiber.

Finally, we shortly mention experiments aiming at using crystals
as a source of photon pairs with embedded memory. In these
protocols, the crystal is illuminated with classical pulses,
creating non classical correlations between an emitted photon and
a stored collective atomic excitation. The protocols include the
rephasing of amplified spontaneous emission (RASE)
\cite{Ledingham2010} and a combination of the DLCZ and AFC scheme
\cite{Sekatski2011}. First demonstrations of the RASE scheme have
been realized, with strong but still classical correlations
obtained in Pr$^{3+}$:Y$_2$SiO$_5$ \cite{Beavan2012} in the photon counting
regime, and evidence of non-classical correlations obtained with
homodyne detection in Tm$^{3+}$:YAG \cite{Ledingham2012}.

\section{Quantum Light sources compatible with solid state quantum memories}
\label{Sec:SPDCsource:General}

In order to achieve strong interactions between a single photon
and a crystal, and to achieve high efficiency storage, it is
crucial that the quantum light has spectral properties that match
those of the quantum memory. The bandwidth of AFC memories is
given by the width of the AFC that can be created in the crystal.
For obtaining the high-finesse combs necessary to achieve high
efficiency storage, the width of the comb is limited by the
spacing between the adjacent states in the ground or excited state
manifolds. Note that AFC broader than the spacing between ground
and excited states can be created, however with a low finesse
leading to limited storage efficiencies, see e.g.
\cite{Saglamyurek2011,Bonarota2011}. In the case where the spacing
becomes bigger than the inhomogeneous broadening of the optical
transition, the limit is then given by the latter. This situation
could be encountered with Kramers ions (e.g. Nd$^{3+}$, Er$^{3+}$), where
moderate magnetic fields could split the states by several GHz .
In principle, high efficiency storage using the AFC echo scheme
could therefore reach GHz bandwidth. In practice however, creating
high quality combs for long storage times over such a large
bandwidth is experimentally challenging. For achieving spin-wave
storage, non-Kramers ions such as Pr and Eu are good candidates.
In these materials, the spacing between hyperfine states are much
smaller, leading to much smaller bandwidths, which can be as low
as a few tens of MHz for Eu$^{3+}$:Y$_2$SiO$_5$  and 4.6 MHz for the excited state
of Pr$^{3+}$:Y$_2$SiO$_5$.  The creation of single photons with such a narrow
linewidth is experimentally challenging. Designing and
implementing  narrowband quantum light sources
\cite{Ou1999,Pomarico2009,Chuu2012,Foertsch2013,Kaiser2013} that
can be interfaced with atoms has been the subject of several
investigations in recent years
\cite{Chaneliere2005,Neergaard-Nielsen2007,Bao2008,Scholz2009,Haase2009,Wolfgramm2011}.
(see also chapters (Chih-Sung Chuu, Mitchell, Zhao)).

The quantum light sources compatible with solid-state quantum
memories that have been realized up to now make use of spontaneous
parametric down conversion (SPDC), where a pump photon is
probabilistically split into a photon pair, with energy and
momentum conservation. A great advantage of this solution is that
it is very flexible in terms of wavelengths of the created
photons, which allows the coupling to quantum memories operating
at any wavelengths. Moreover, it can create two photons with
different frequencies, which can be used for example to create non
classical correlations between a quantum memory operating in the
visible range and a photon at telecommunication wavelengths.
However, the spectrum of the photons emitted by spontaneous down
conversion typically goes from 100 GHz to THz, several orders of
magnitude larger than the bandwidth of quantum memories. Therefore
extensive filtering must be applied in order to generate quantum
memory
compatible quantum light using spontaneous down conversion. \\

Filtering can be applied after the source, using passive filters.
However this requires extremely bright sources \cite{Clausen2011},
e.g. waveguide sources. The waveguide increases the production
rate of photon pairs significantly \cite{Tanzilli2001}, allowing
for an efficient source while pumping it with a low peak-power cw
laser. This is particularly important for a strongly filtered
source, in order to have a sufficiently high probability of
creating a photon pair within the filtered spectral regions. It is
important to note that when passive filtering is used, the number
of photon pairs per coherence time (or the spectral brightness
expressed in pairs per second per mW of pump power and per MHz of
bandwidth) does not change with the filter width
\cite{Halder2008}. While the rate of created pairs per second
decreases with the filter width, the coherence time of the photon
increases, leading to a constant spectral brightness.

Another way of implementing a narrow-band quantum light source
from spontaneous down conversion is to insert the non linear
crystal in an optical cavity \cite{Ou1999} (see also chapters
(Zhao,Mitchell)). This not only has the advantage of providing
filtering, but also enhances the probability of generating a
photon within a cavity mode, with respect to the no cavity case.
In the ideal case, the enhancement is given by $Q  = F^3/(\pi
F_0)$ \cite{Ou1999}, where $F$ is the finesse of the cavity and
$F_0$ is the finesse calculated only from the mirrors
reflectivity. In order to reach this enhancement, both signal and
idler fields need to be resonant with the cavity.

Beyond the use of SPDC, several other systems could be used as
quantum light sources. It could be possible to use the doped
crystals themselves as sources of photon pairs with the required
spectral properties, as shown with quantum memories based on
atomic gases (see. e.g. \cite{Sangouard2011} for a review).
However, this turns out to be much more difficult to implement in
rare-earth doped solids, due to the very small oscillator strength
of the optical transition. Several schemes have been proposed,
including the rephasing of amplified spontaneous emission
(RASE)\cite{Ledingham2010} and a combination of the DLCZ and AFC
scheme \cite{Sekatski2011}. Solid state single photon emitters may
also be used as compatible single photon sources, e.g. quantum
dots or single molecules. Quantum dots have typically GHz
spectral bandwidths (see chapters Lanco, Schneider,McMahon), which may be compatible with the AFC
echo scheme with broadband combs. The challenge is to tune the
quantum dot in resonance with the rare-earth ensemble, without
broadening the line. Single molecules in solid state matrices have
shown much narrower spectral bandwidth, down to tens of MHz
\cite{Lettow2010}, which could be used with spin-wave quantum
memories. Again, the challenge is to tune them near resonance. A
potential solution for the frequency mismatch is to implement
quantum frequency conversion \cite{Tanzilli2005} which has been
used recently to interface telecom photons to quantum memories
\cite{Albrecht2014,Maring2014}.

\subsection{Characterizing photon pair
sources}

In this section, we discuss various ways to characterize photons
pairs emitted by SPDC and to quantify the correlations between
signal and idler fields. The state created by single mode SPDC is
given by (for $p\ll 1$):

\begin{equation}
\label{Eq:TMSS}
|\Psi\rangle_{s,i}=\sqrt{1-p}\sum_{n=0}^{\infty}p^{\frac{n}{2}}|n\rangle_s|n\rangle_i
\end{equation}
where $p$ is the probability to create a photon pair and is
proportional to the pump power, and $|n\rangle_s (|n\rangle_i)$ is
a n photon Fock state in the signal (idler) mode.
 This state is known as two-mode
squeezed state. It displays very strong quantum correlations
between the two modes, i.e. the signal and idler fields.

For a photon pair source, the quality of the correlations between
signal and idler fields is usually quantified measuring the second
order cross-correlation function $G_{s,i}^{(2)}(\tau)$ between the
two fields, by performing a coincidence measurement. The
normalized form of $G_{s,i}^{(2)}(\tau)$, denoted as
$g_{s,i}^{(2)}(\tau)$ can be expressed as:

\begin{equation}
g_{s,i}^{(2)}(\tau) \equiv \frac{\langle E_s^{\dagger}(t)
E_i^{\dagger}(t+\tau) E_i(t+\tau) E_s(t) \rangle}{\langle
E_i^{\dagger}(t+\tau) E_i(t+\tau) \rangle
\langle E_s^{\dagger}(t) E_s(t) \rangle}, \\
\end{equation}
where $E_{s,i}^{\dagger}$ ($E_{s,i}$) is the electric field
creation (annihilation) operator for the signal and idler fields.
For the ideal two mode squeezed state of Eq. \ref{Eq:TMSS}, the
cross-correlation function is given by :
\begin{equation}
g^{(2)}_{s,i}=1+\frac{1}{p}
\end{equation}
We see that the correlation decreases when increasing $p$. This is
the consequence of the creation of multiple pairs. We also see
that $g^{(2)}_{s,i}$ can become arbitrarily high for low
excitation probability. However, this comes at the expense of the
count rate, since the mean number of photons in the signal mode is
given by
\begin{equation}
 \bar{n}_s=\frac{p}{1-p}
\end{equation}
This illustrates a fundamental limitation of SPDC: there is a
trade off between the degree of quantum correlation and the count
rate that can be obtained. In practice, $g^{(2)}_{s,i}$ is
determined over a detection
 window
$\Delta \tau$ as :
\begin{equation}
g^{(2)}_{s,i}(\Delta \tau)=\frac{p_{s,i}}{p_sp_{i}}
\end{equation}
where $p_{s,i}$ is the probability to detect a coincidence between
signal and idler photons and $p_s$ ($p_{i}$) is the probability to
detect a signal (idler) photon in a time interval $\Delta \tau$.

The second order cross-correlation function can also be used to
gain information on the spectral content of the photon pairs. For
example, in the case of doubly-resonant cavity-enhanced
downconversion, following the theory used in
\cite{Scholz2009,Wolfgramm2011},  $g_{s,i}^{(2)}(\tau)$ takes the
form:
\begin{eqnarray}
    \begin{split}
        g_{s,i}^{(2)}(\tau) \propto {} & \Bigg| \sum_{m_s, m_i = 0}^\infty \frac{\sqrt{\gamma_s \, \gamma_i \, \omega_s \, \omega_i}}{\Gamma_s + \Gamma_i} \\
                        & \times
                      \begin{cases}
                                e^{-2 \pi \Gamma_s (\tau-(\tau_0/2))}{\rm sinc}{(i \pi \tau_0 \Gamma_s)} \hspace{3 mm} \hspace{0.5 mm} \tau \geq \frac{\tau_0}{2}\\
                                e^{+2 \pi \Gamma_i (\tau-(\tau_0/2))}{\rm sinc}{(i \pi \tau_0 \Gamma_i)} \hspace{4 mm} \hspace{0.5 mm} \tau < \frac{\tau_0}{2}
                      \end{cases} \hspace{-4 mm} \Bigg|^2,
    \end{split}
    \label{eq:G2}
\end{eqnarray}
where $\gamma_{s,i}$ are the cavity damping rates for signal and
idler, $\omega_{s,i}$ are the central frequencies,
 $\Gamma_{s,i} = \gamma_{s,i}/2+i m_{s,i} \rm{FSR}_{s,i}$ with mode indices $m_{s,i}$ and
free spectral ranges $\rm{FSR}_{s,i}$, and $\tau_0$ is the transit
time difference between the signal and idler photons through the
SPDC crystal. Eq. \ref{eq:G2} shows that the second order
cross-correlation function for a multimode cavity output displays
an oscillatory behavior with peaks separated by the inverse of the
cavity free spectral range. The width of the peaks is directly
related to the number of spectral modes in the signal and idler
fields. For two modes, the oscillation will be sinusoidal, and the
width will then decrease with the number of modes. In practice,
the minimum width that can be detected is given by the detectors'
time resolution.

For the single mode case, we have :

\begin{equation}
g_{s,i}^{(2)}(\tau)=1+\frac{4}{p}\frac{\gamma_s\gamma_i}{(\gamma_s+\gamma_i)^2}f(\tau)
\end{equation}
where
\begin{eqnarray}
\begin{split}
f(\tau)=
\begin{cases}
e^{-\gamma_s \tau} \hspace{3 mm} \hspace{0.5 mm} \tau \geq 0\\
e^{\gamma_i \tau} \hspace{4 mm} \hspace{0.5 mm} \tau < 0
\end{cases}
\end{split}
\end{eqnarray}

This is the same expression obtained for the case without cavity,
but with single mode Lorentzian filters with width $\gamma_s$ and
$\gamma_{i}$ inserted in the signal and idler modes, respectively
\cite{Clausen2014a}.

Information about the spectral content of the created photons can
also be obtained by measuring the unconditional second order
autocorrelation function $g^{(2)}_{s,s}$ and $g^{(2)}_{i,i}$, for
the signal and idler fields, respectively. For an ideal two mode
squeezed states, the unconditional field exhibit thermal
statistics with $g^{(2)}_{s,s}(0)= g^{(2)}_{i,i}(0)=2$. However,
if several spectral modes are present, it has been shown
\cite{Christ2011} that the autocorrelations decrease as
$g^{(2)}_{s,s}(0)= g^{(2)}_{i,i}(0)=1+1/K$ where $K$ is the number
of modes. By measuring the unconditional second order
autocorrelation function, one can therefore bound the number of
spectral modes present in the fields. This is valid as long as the
measurement is not limited by noise.

The non-classical nature of the correlation between signal and
idler fields can be experimentally assessed with a Cauchy-Schwarz
inequality. For a pair of independent classical fields the
following inequality must be fulfilled:
\begin{equation}
R=\frac{(g^{(2)}_{s,i})^2}{g^{(2)}_{s,s}g^{(2)}_{i,i}}\leq1
\end{equation}
If the signal and idler fields exhibit thermal or sub-thermal
statistics ($g^{(2)}_{s,s}(0)= g^{(2)}_{i,i}(0)\leq 2$, the
measurement of $g^{(2)}_{s,i}>2$ is therefore a signature of
non-classical correlations. However, in order to prove
non-classicality without assumptions on the created state, the
unconditional autocorrelation functions should be measured as
well.

\subsection{A quantum light source compatible with Nd doped crystals}
\label{Sec:SPDCsource:Nd}

\begin{figure}[b]
    \includegraphics[scale=.40]{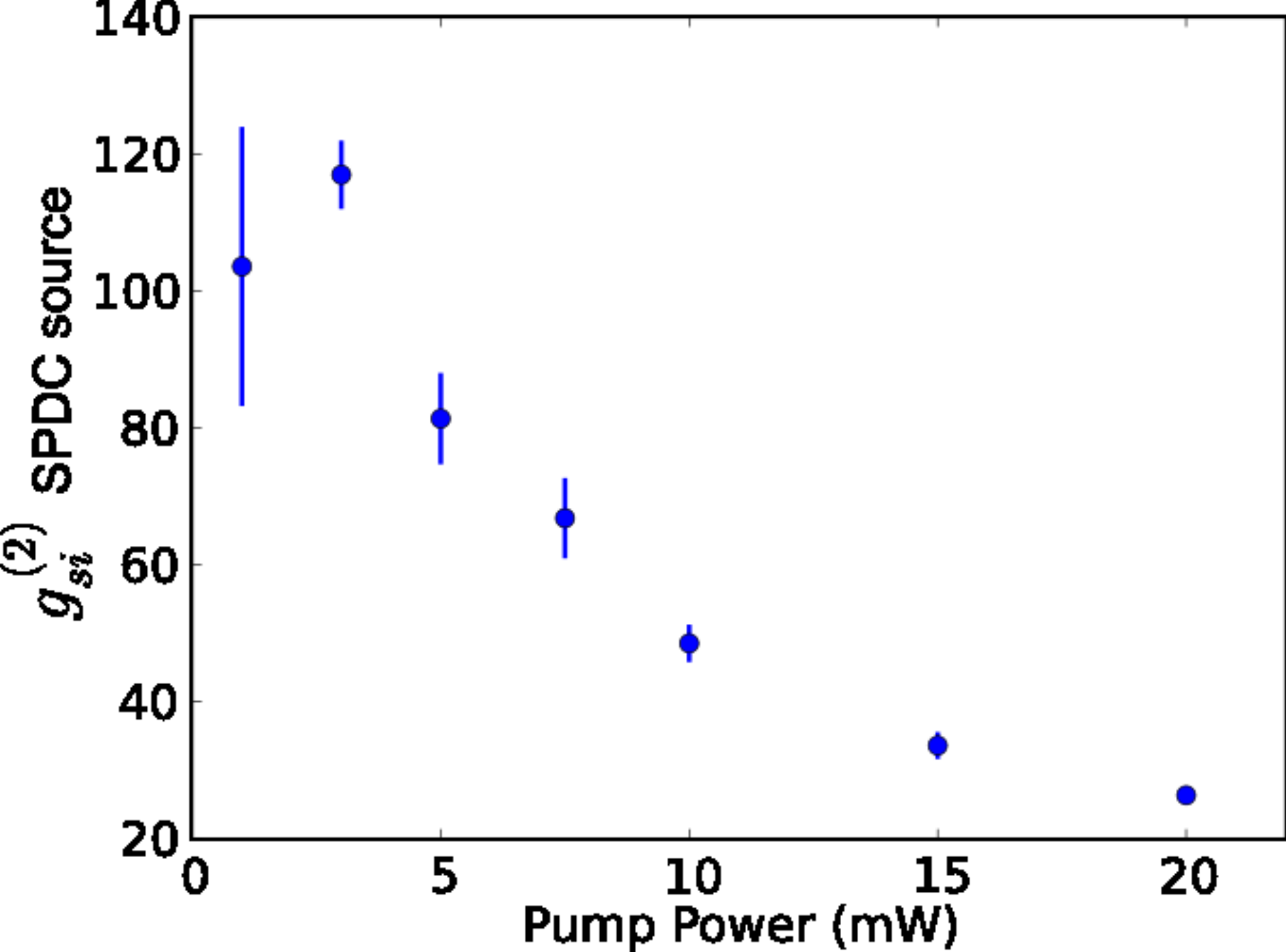}
    \sidecaption[t]
    \caption{Second-order cross-correlation $g^{(2)}_{s,i}(\Delta \tau)$
function of the filtered SPDC source developed for Nd doped
crystals, as a function of the power of the laser pumping the SPDC
source. The detection integration window was $\Delta \tau$ = 10
ns.}
    \label{fig:source_Nd_corr_funcs}
\end{figure}

Here we discuss a filtered SPDC source producing photons at 883 nm
(signal) and 1338 nm (idler), which was developed to interface
with a Nd$^{3+}$:Y$_2$SiO$_5$ quantum memory operating at 883 nm,
having a memory bandwidth of 120 MHz. This particular source has
been described in Refs \cite{Clausen2011,Usmani2012}, while a
similar source, slightly more broadband, of polarization-entangled
photons for quantum storage was described in a more recent work
\cite{Clausen2014a}.

The SPDC source was based on a periodically-poled potassium
titanyl phosphate (PPKTP) crystal with an optical waveguide. The
PPKTP crystal was pumped by a continuous-wave (cw) 532 nm laser,
which is convenient since powerful and frequency stable
single-mode Nd:YAG lasers exist at this wavelength. Since one
photon should be resonant with the 883 nm transition in
Nd$^{3+}$:Y$_2$SiO$_5$, the choice of the pump laser imposed the
wavelength of the idler photon to 1338 nm, in the
telecommunication O-band. In principle one could use another pump
wavelength to produce an idler photon in the more conventional
telecommunication C-band around 1550 nm.

Without external frequency filtering the source produced photons
with a bandwidth of about 800 GHz, such that strong filtering of
the photon pairs was necessary to match the memory bandwidth of
120 MHz. To this end a combination of diffraction gratings,
optical cavities and a fibre-based filter was used. The gratings
provided full-width at half-maximum (FWHM) bandwidths of 90 and 60
GHz for the signal and idler photons, respectively. For the signal
photon two etalons placed in series resulted in a single
longitudinal mode with a bandwidth of 350 MHz. It should also be
emphasized that the 6 GHz wide inhomogeneous absorption profile of
the Nd$^{3+}$:Y$_2$SiO$_5$ crystal provided additional filtering,
since the 120 MHz comb was created within this absorption profile.
For the idler photon a home-made narrow-band cavity filtered down
the photons to a FWHM linewidth of 43 MHz. A single longitudinal
mode of this cavity was selected by a fiber Bragg grating (FBG).
The total transmission coefficients from the PPKTP waveguide to
the single-mode fibers were 22\% and 14\% for signal and idler
photons \cite{Usmani2012}, respectively, including fiber coupling.
The more recent version of this source \cite{Clausen2014a} reach
higher efficiencies, partly because high-efficiency volume Bragg
gratings (VBGs) replaced the diffraction gratings.

Strong non-classical correlations between the signal and idler photons can only be obtained if the central frequencies of the filters on each mode satisfy the energy conservation of the SPDC processes. This is a non-trivial task when dealing with highly non-degenerate SPDC sources, particularly when one mode must be resonant with an external quantum memory. A solution to this problem was introduced in the work discussed here. A reference laser that is resonant with the signal filtering system and the quantum memory is injected into the SPDC source. This will create light at the idler wavelength through difference frequency generation (DFG), which obey the necessary energy conservation. This can be used to adjust central frequency of the idler filtering system, alternatively one can also change the wavelength of the pump laser. 

The filtered SPDC source can be characterized by the pair creation
rate within the filtered modes (spectral brightness) and the
second-order auto and cross correlation functions of the signal
and idler modes. The intrinsic spectral brightness of the PPKTP
waveguide was 6.3$\cdot$10$^3$ pairs/(mW $\cdot$ MHz $\cdot$ s)
\cite{Usmani2012}, which does not include the transmission through
the filtering elements. This means that the probability $p$ to
create a pair (in the limit $p \ll 1$) is then $p \approx 2.7
\cdot 10^{-3}$/mW per 10 ns within a 43 MHz wide spectral window.
The duration of the detection window should be set with respect to
the photon pair coherence time, which in our case is dominated by
the 43 MHz filter on the idler side (7 ns coherence time). The
spectral brightness of the source including the filtering elements
and fiber coupling was $\approx$ 200 pairs/(mW $\cdot$ MHz $\cdot$
s). If we again assume a 43 MHz filtered bandwidth we arrive at a
final pair creation rate of 8600 pairs/s per mW pump power.

The second-order auto-correlation functions of both the idler and
signal modes were characterized \cite{Usmani2012}, resulting in
$g_{i,i}^{(2)}(0)=1.9$ and $g_{s,s}^{(2)}(0)=1.8$. These values
give effective mode numbers $K=1.1$ and $K=1.25$ for the idler and
signal modes, which indicate single frequency modes on both the
signal and idler side.

In Figure \ref{fig:source_Nd_corr_funcs} we show the second-order
cross-correlation function $g_{s,i}^{(2)}$ of the source as a
function of pump power which exhibits the expected $1+1/p$
behaviour \cite{Clausen2011}. Based on the intrinsic spectral
brightness we expect a cross-correlation $g_{s,i}^{(2)} \approx
75$ for $P=5$ mW, which is very close to the measured value of
about 80. The cross-correlation function is larger than the
classical upper bound of 2, where one assumes
$g_{i,i}^{(2)}(0)=g_{s,s}^{(2)}(0)=2$, for all pump powers,
clearly demonstrating the strong non-classical correlations of the
source.

\subsection{A quantum light source compatible with Pr doped
crystals} \label{Sec:SPDCsource:Pr} In the context of generating
quantum light compatible with Pr$^{3+}$:Y$_2$SO$_5$, one of the
photons must be at 606 nm, with a bandwidth smaller than 4 MHz. In
addition, in order to use this source to generate entanglement
between remote crystals, it is desirable to have the second photon
of the pair at a telecom wavelength. An experiment showing these
properties was reported in \cite{Fekete2013}.

\begin{figure}[h]
    \includegraphics[scale=.45]{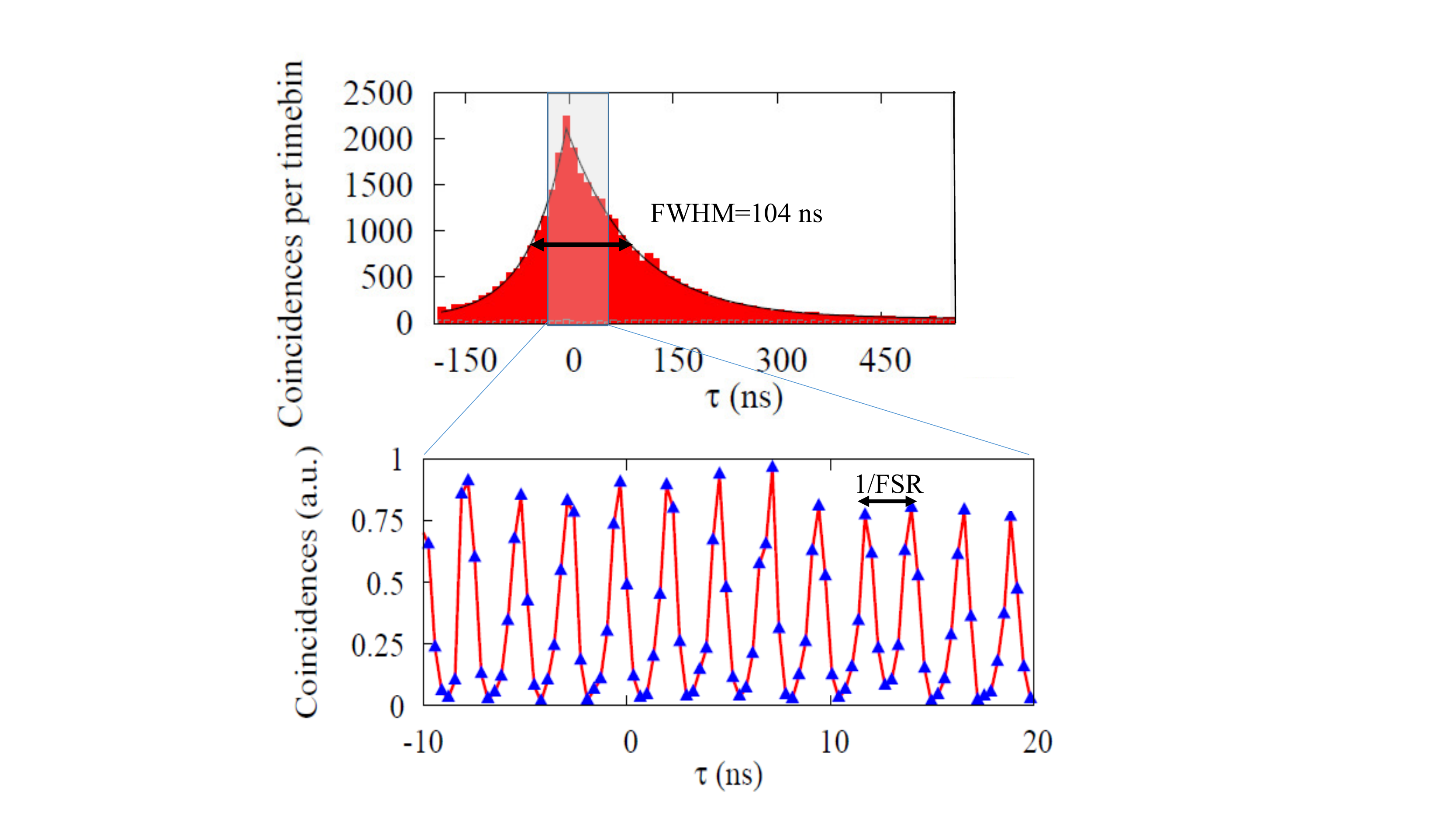}
    \caption{Measured second-order cross-correlation $G^{(2)}_{s,i}(\tau)$
function (non normalized) of the cavity enhanced SPDC source
developed for Pr doped crystals \cite{Fekete2013}. The FWHM correlation time is 104
ns. The zoom displays a higher temporal resolution, allowing to
observe the oscillations in the $G^{(2)}_{s,i}(\tau)$,
characteristic of multi spectral mode output. From this data, a
number of four mode per cluster is inferred.}
    \label{fig:source_Pr}
\end{figure}

In order to meet these requirements, the following source was
used. A pump laser at 426 nm was used to generate widely non
degenerate photon pairs at 606 nm and 1436 nm in a periodically
poled lithium niobate  crystal. The crystal was placed in a bow
tie optical cavity with a free spectral range of 400 MHz and a
finesse of around 200 for the signal and idler modes. A special
coating was used to achieve high reflectivity (99.99 $\%$) for
three mirrors, and a specified reflectivity of 98.5 $\%$ for the
output coupler. The SPDC output is modified by the cavity modes.
In the case of degenerate photons with the same polarization, the
spectrum would be a convolution between the phase matching
bandwidth and the cavity modes. However, in the non degenerate
case, the signal and idler modes experience different dispersion
characteristics in the crystal, which leads to slightly different
free spectral ranges. Since both modes have to be resonant with
the cavity to enable enhancement, only a subset of modes will be
enhanced. These modes are grouped in so called clusters
\cite{Pomarico2009,Pomarico2012}. The width and spacing of the
clusters is determined by the dispersion properties of the
crystals and the cavity geometry.

In the experiment of Ref. \cite{Fekete2013}, the number of modes
per cluster was inferred to be around 4, by looking at the
oscillations in the  $g_{s,i}^{(2)}(\tau)$ function and comparing
to Eq. \ref{eq:G2} (see Fig. \ref{fig:source_Pr}). This number was
also verified by direct measurement of the signal and idler fields
with a narrow band filter cavity, scanned over the spectrum. The
number of clusters was then inferred by measuring the first order
correlation function of the idler field. Within the phase-matching
bandwidth of the free space down conversion photons of around
$\Gamma_{pm}$= 80 GHz, the spectrum of the photons leaving the
cavity was finally inferred to be composed of one main cluster
containing around four longitudinal modes, and two smaller side
clusters separated by 45 GHz from the main one and suppressed by
around 80 $\%$. For degenerate photons, the number of longitudinal
modes would be given by $\Gamma_{pm}/FSR$ = 200 modes. The
clustering effect therefore leads to a suppression of the number
of modes of around 50. Within a longitudinal mode, the photons
pair created featured a correlation time of 104 ns (see Fig.
\ref{fig:source_Pr}), the longest demonstrated so far with SPDC
sources. From the decay of the $G^{(2)}_{s,i}(\tau)$ (fitted with
exp($-2\pi\Delta\nu\tau$)), spectral-linewidth of $\Delta \nu$ of
1.7 MHz and 2.9 MHz were inferred for the idler field at 1436 nm
and the signal field at 606 nm, respectively. The asymmetry can be
attributed to the different intra-cavity losses for the two
wavelengths. A intrinsic spectral brightness of 8$\cdot$10$^3$
pairs/(mW $\cdot$ MHz $\cdot$ s) was inferred, leading to $p
\approx 6.4 \cdot 10^{-3}$/mW per 400 ns within a 2 MHz wide
spectral window. However due to large optical losses, the spectral
brightness before the detectors (inside single mode fibers) was 11
pairs/(mW $\cdot$ MHz $\cdot$ s). In a more recent version of the
source, this number was increased to 190 pairs/(mW $\cdot$ MHz
$\cdot$ s).

This source still requires additional filtering for selecting a
single frequency mode. However, the filtering requirement are
considerably relaxed due to the low number of modes present in the
spectrum. Side clusters can easily be removed by placing etalons
with high transmission efficiency in the signal or idler modes. In
order to select a single mode, a narrow band filter cavity with
FSR = 16.8 GHz and linewidth of 80 MHz has been placed in the
signal arm. In this configuration, the measured
$G_{s,i}^{(2)}(\tau)$ was measured and no oscillation was
observed, confirming that only one mode per cluster was present.
The suppression of the side cluster could also be inferred first
order autocorrelation.

\section{Quantum Light Storage experiments}
\label{Sec:QStorage}
\subsection{Quantum entanglement storage in Nd:YSO crystals}
\label{Sec:QStorage:Nd} 
\begin{figure}[th]
\includegraphics[scale=.50]{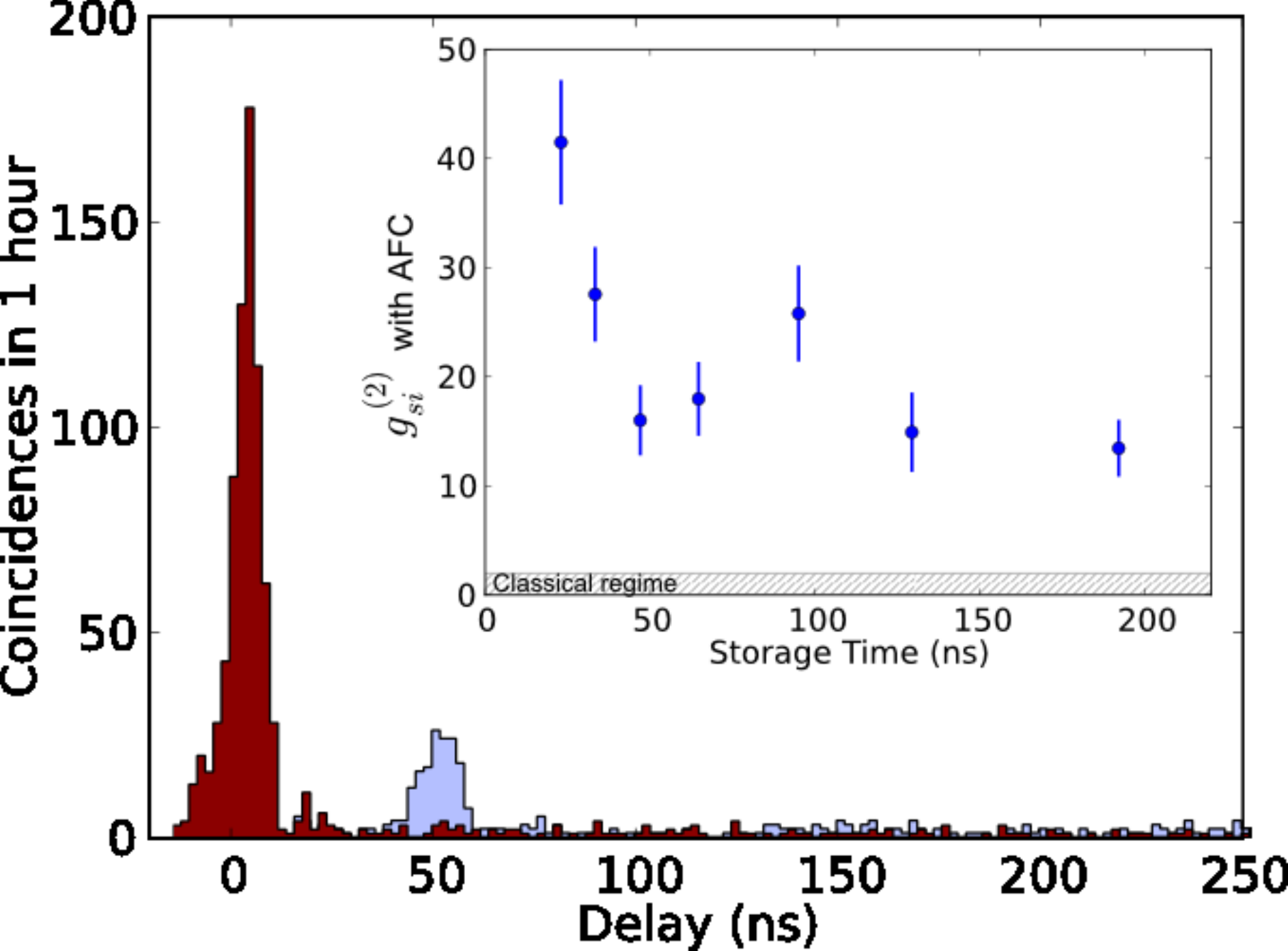}
 \caption{Detection event histogram as a function
of time. The peak at around zero time was recorded by making a
large transparency hole in the Nd$^{3+}$:Y$_2$SiO$_5$ absorption,
therefore it represents the input mode. When creating an AFC with
periodicity $\Delta$ = 20 MHz, the expected AFC echo appears as a
peak at around 50 ns. The excellent signal-to-noise ratio
indicates a strong quantum correlation between the idler photon
(1338 nm) and the signal photon (883 nm), after the latter has
been stored in the memory. The inset quantifies this by showing
the second-order cross-correlation $g^{(2)}_{s,i}(\Delta \tau)$
function as a function of storage time $1\Delta$. The
$g^{(2)}_{s,i}(\Delta \tau)$ function is significantly larger than
the classical limit of 2, for all storage times. The detection
integration window was $\Delta \tau$ = 10 ns.}
\label{fig_histNd}       %
\end{figure}

In 2008 the first AFC echo storage
experiment in a Nd$^{3+}$-doped YVO$_4$ crystal at the single
photon level was demonstrated \cite{Riedmatten2008}, which was the
starting point for considering storing true quantum states of
light. To this end the filtered SPDC source described in section
\ref{Sec:SPDCsource:Nd} was developed. The first Nd:YVO$_4$ memory
had a bandwidth of a few MHz, making the filtering of the SPDC
source difficult. A more wideband quantum memory was therefore
developed in a Nd$^{3+}$-doped crystal Y$_2$SiO$_5$, using the
$^4$I$_{9/2}-^4$F$_{3/2}$ transition at 883 nm
\cite{Usmani2010,Clausen2011}. This optical transition has an
inhomogeneous broadening of about 6 GHz. The ground state was
split into a Kramers spin doublet by applying a magnetic field of
around 300 mTesla, producing a Zeeman split of about 11 GHz. The
comb structure was created on one of the Zeeman transitions by
performing spectral hole burning, i.e. ions were optically pumped
into the other Zeeman spin state. In practice this was done by
scanning the laser frequency of a narrow-band external-cavity
diode laser (EDCL) with a acousto-optic modulator (AOM), while
periodically switching off and on the light. The total bandwidth
was 120 MHz, limited by the scan range of the AOM. The maximum
efficiency at low storage times $1/\Delta$=25 ns was 20\%, close
to the optimal value for the optical depth of the crystal. In
Figure \ref{fig_histNd} we show an example of storage of 883 nm
signal photons produced by the filtered SPDC source in the crystal
\cite{Clausen2011}. The cross correlation function between the
signal and idler modes after storing the signal mode clearly shows
quantum correlations, for all storage times. This can be
interpreted as storing a single photon at 883 nm, heralded by the
detection of an idler photon.

The second-order cross-correlation function $g^{(2)}_{s,i}(\Delta
\tau)$ after storage of the signal photon is lower than that of
the source, cf. Figure \ref{fig:source_Nd_corr_funcs}. In
\cite{Usmani2012} it was shown that it was due to the delay
introduced by the memory. Indeed, the signal photons released from
the memory are superimposed with uncorrelated signal photons
created later during the AFC echo emission, which are transmitted
through the memory with a certain probability. This noise source
depends on the ratio of the memory transmission to the memory
efficiency \cite{Usmani2012}. A solution to this problem is to
turn off the SPDC pump laser before the AFC echo, which was done
in the quantum storage experiment in Pr$^{3+}$:Y$_2$SiO$_5$ described below
\cite{Rielander2014}.

The photons produced by a SPDC source pumped by a CW pump laser
are entangled in energy-time \cite{Franson1989}, provided that the
coherence time of the pump is significantly longer than the
coherence time of the signal-idler pair. This condition is
naturally met when using a single-mode frequency CW pump laser, as
in this case. This energy-time entanglement can be revealed by
making projective measurements on different time basis states, on
both the signal and idler modes. Practically, one can place
Mach-Zehnder (MZ) interferometers in each mode, each MZ having the
same path difference $\Delta T$, which is known as a Franson-type
set-up \cite{Franson1989}. The two-photon interference fringes
obtained when varying the phases in each interferometer reveals
the energy-time entanglement. The coincidence rate varies like
$V\cos(\Delta \phi_s+\Delta \phi_s)$, where $\phi_s$ and $\phi_i$
are the phase settings on the signal and idler interferometers and
$V$ is the visibility. More formally the quantum entanglement can
be detected by a violation of the Clauser-Horne-Shimony-Holt
(CHSH) inequality \cite{Clauser1969}, where the CHSH parameter $S$
is larger than 2 for any entangled state. The presence of
entanglement can also be inferred from a fringe visibility larger
than $1/\sqrt{2} \approx 70.7\%$.

\begin{figure}[t]
    \includegraphics[scale=.6]{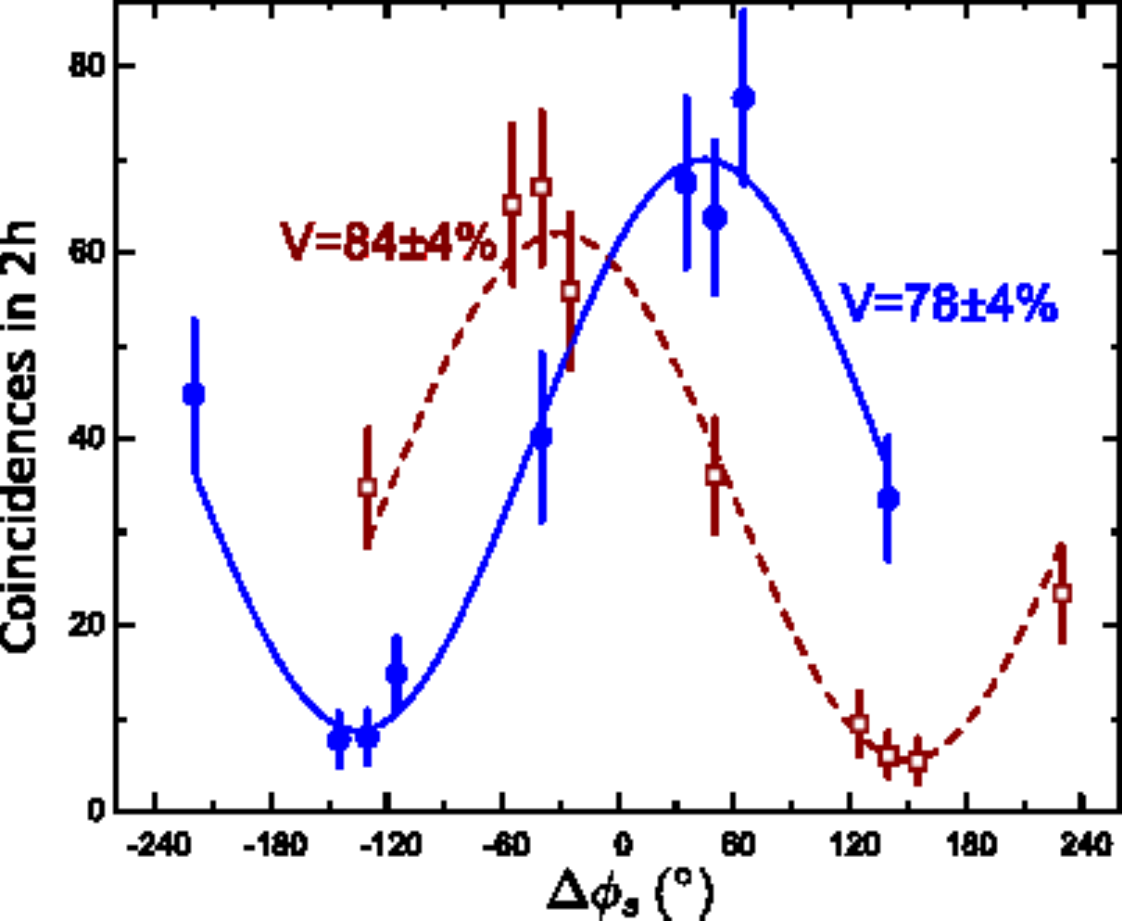}
    \sidecaption[t]
    \caption{Number of detected signal-idler coincidences as a function of the signal analyser setting $\Delta \phi_s$, for two settings $\Delta \phi_i$ of the idler analyser shown as square and circle symbols. The solid ($V=78 \pm 4 \%$) and dashed ($V=84 \pm 4 \%$) lines are fits to the circles and squares, respectively.}
    \label{fig:energy-time-fringes}
\end{figure}

In the experiment we describe here \cite{Clausen2011}, a
particular twist was introduced. While the idler photon was
analyzed using a standard interferometer, the signal photon was
analyzed inside the Nd$^{3+}$:Y$_2$SiO$_5$ memory. We exploited
the fact that more complicated absorption features than periodic
combs can be created. Indeed, by creating an absorption structure
that was the sum of two combs with different periodicities
$\Delta_1$ and $\Delta_2$, we could create an effective unbalanced
interferometer within the crystal. By setting
$1/\Delta_1-1/\Delta_2 = \Delta T$ we could analyze the
entanglement \textit{in memory}.

In Figure \ref{fig:energy-time-fringes} we show examples of
interference fringes as a function of $\Delta \phi_s$ for two
values of $\Delta \phi_i$. Both visibilities are well above the
limit $\approx 70.7\%$, strongly indicating the presence of
entanglement. By explicitly measuring a CHSH parameter of $S =
2.64 \pm 0.23$ the entanglement between the idler photon and the
photon stored in the crystal was clearly demonstrated.

In parallel to the work described above, Saglamyurek \textit{et
al.} also demonstrated storage of a photon entangled with a
another photon stored in and released from a rare-earth doped
crystal \cite{Saglamyurek2011}. In their experiment the entangled
photons were produced by a bulk periodically poled lithium niobate
(PPLN) crystal, which was pumped using a pico-second laser with
high peak power. Time-bin entangled photons were produced by
splitting the 16 ps pump pulse into two coherent pulses separated
by 1.4 ns. The idler photon was at the telecom wavelength 1532 nm,
while the signal photon was at 795 nm. The quantum memory was
based on a atomic frequency comb created on the
$^3$H$_6$-$^3$H$_4$ transition at 795 nm in a Thulium-doped
lithium niobate crystal (Tm$^{3+}$:LiNbO$_3$). A particular
feature of this experiment was the optical waveguide on the
surface of the Tm$^{3+}$:LiNbO$_3$ chip. The waveguide technology
is widespread in integrated optics and opens up interesting
perspectives of combining the storage device with other optical
elements. The AFC memory in this experiment was also very
broadband, the AFC spanned a range of 5 GHz. This allowed storage
of the short 16 ps photons produced by the filtered SPDC source.
This also implies that the degree of filtering was less stringent
than in the experiment described previously. A drawback of this
material is, however, the short storage time and the low
efficiency. The entanglement experiment discussed here
demonstrated a storage time of 7 ns, with an AFC echo efficiency
of 2\%. The preservation of entanglement was proved by testing the
Clauser-Horne-Shimony-Holt(CHSH) inequality, resulting in a CHSH
parameter of S=2.25 $\pm$ 0.06.

In the next experiment that we will describe the goal was to
entangle two Nd$^{3+}$:Y$_2$SiO$_5$ crystals \cite{Usmani2012}.
This could be done by storing each photon out of an entangled
pair, but this requires that both photons are resonant with a
memory. With only one photon in resonance with the Nd doped
crystal at 883 nm, the choice was made instead to store a
path-entangled state of a single photon \cite{Enk2005}. More
specifically, this state can be created by sending a single photon
state $|1\rangle$ through a balanced beam-splitter, which creates
the state
$1/\sqrt{2}(|1\rangle_A|0\rangle_B+|0\rangle_A|1\rangle_B)$ of the
spatial output modes A and B of the beam-splitter. This state is
sometimes referred to as single-photon entanglement
\cite{Enk2005}. Now, entanglement between two memories can be
realized by placing one memory in each path. This approach was
first used by Choi \textit{et al.} \cite{Choi2008} to entangle two
spatial modes in the same cloud of laser-cooled caesium atoms,
where each mode was stored in the cloud using EIT. In the
experiment described here, this this approach was used to entangle
the modes of two physically distinct crystals and the modes where
stored using the AFC scheme \cite{Usmani2012}. The single photon
state was produced by detecting the idler photon from the filtered
SPDC source, which creates a state very close to a single photon
in the signal mode (a heralded single-photon source).

To characterize the single-photon entanglement one can read out
the memories and then performing measurements on the photonic
state, which provides a lower bound of the entanglement present
while storing the two modes. To detect the entanglement we used
the tomographic approach developed by Chou \textit{et al.}
\cite{Chou2005}, in which the entanglement is quantified through a
single parameter, the concurrence $C$. The concurrence is positive
$C>0$ for an entangled state, more specifically a separable state
gives $C=0$ and a maximally entangled state $C=1$. To compute the
concurrence one needs to measure the probability of finding
exactly one signal photon in any of the two modes, $p_{01}$ and
$p_{10}$, and the probability of finding one signal photon in each
path, $p_{11}$. This requires two- and three-fold coincidence
measurements, respectively, since one also needs to detect the
heralding idler photon. In addition one needs to measure the
one-photon visibility $V$ associated with the two paths A and B.
The concurrence can then be calculated with the formula:
$C=V(p_{10}+p_{01})-\sqrt{2p_{00}p_{11}}$ \cite{Chou2005}. By
performing all the necessary measurements for a relatively high
pump power of 16 mW, we reached a concurrence of $C=6.3 \pm 3.8
\cdot 10^{-5}$ indicating presence of entanglement
\cite{Usmani2012}. The low concurrence was essentially due to
losses, since the detection probabilities include all propagation
losses, memory efficiency and detection probabilities.
Alternatively one can use an approach where the $p_{11}$
probability is inferred from the cross-correlation function
$g^{(2)}_{s,i}$, which is a much more time efficient experiment as
it uses only two-fold detections. Using this approach the presence
of entanglement at the pump power of 16 mW could be confirmed, and
the concurrence was measured for a range of lower pump powers
where the direct method turned out to be too time consuming due to
the rarity of three-fold coincidences. The concurrence reached
about $C=1.1 \pm 0.1 \cdot 10^{-4}$ for the lowest pump power of 1
mW, showing the presence of entanglement of the stored modes in
the two Nd$^{3+}$:Y$_2$SiO$_5$ crystals.

\subsection{Quantum Storage of heralded single photon in a Pr$^{3+}$:Y$_2$SiO$_5$ crystal}
\label{Sec:QStorage:Pr} As mentioned previously, Pr doped solids
have demonstrated exceptional properties for light storage
experiments, including long storage times
\cite{Longdell2005,Lovric2013,Heinze2013} and high efficiencies
\cite{Hedges2010,Sabooni2013}.
Despite these very promising properties, there is currently only
one demonstration of storage of quantum light in this system. In
this section, we describe in more detail this demonstration,
initially reported in \cite{Rielander2014}.

\begin{figure}[t]

\includegraphics[scale=.45]{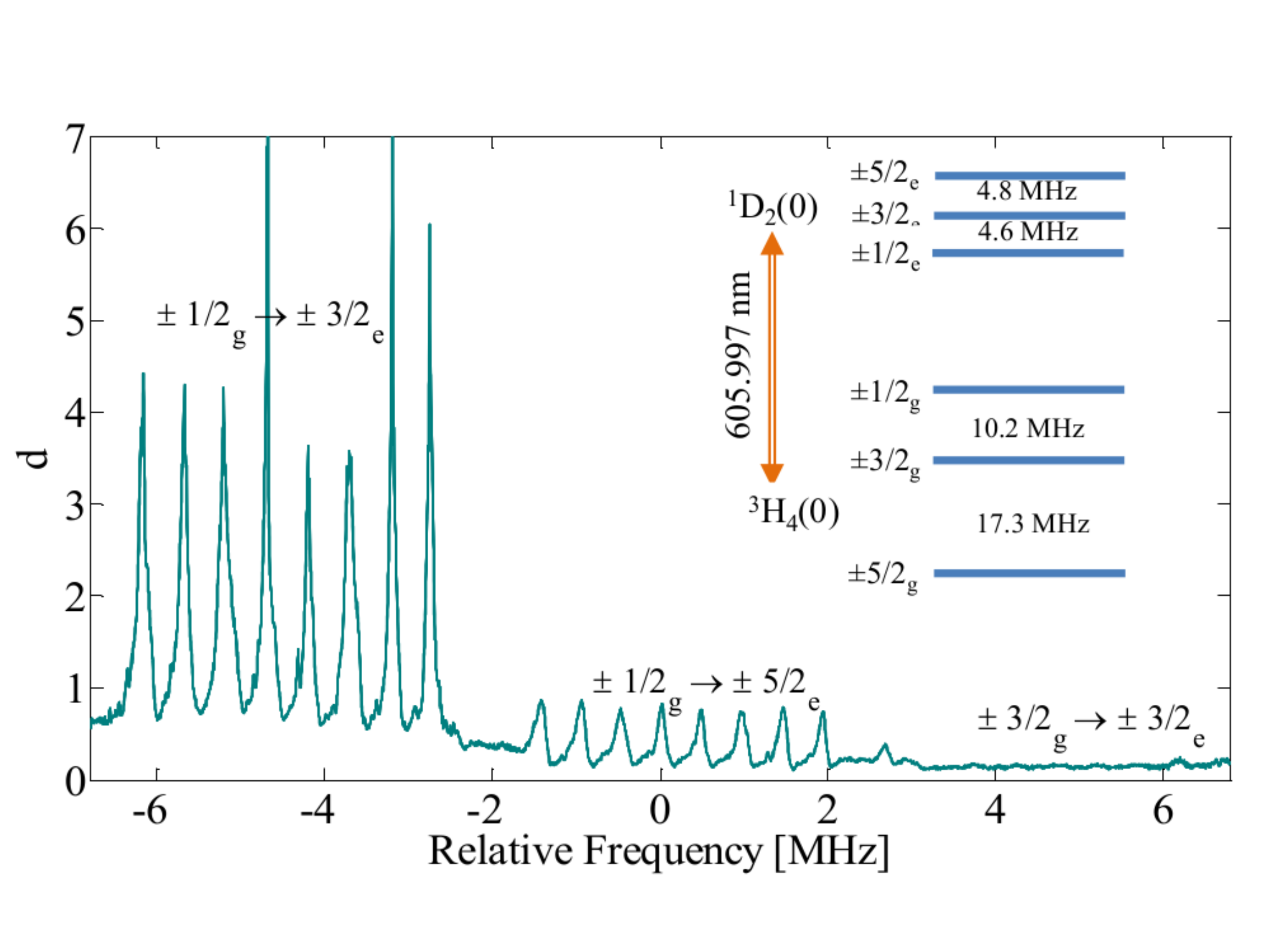}

\caption{ Exemple of an atomic frequency comb in  Pr$^{3+}$:Y$_2$SiO$_5$ . The optical depth ($d$) is plotted as a function of the relative frequency. The input and output photons are resonant with the $\pm 1/2_g$ $\rightarrow$ $\pm 3/2_e$ transition.  Inset : Relevant energy level scheme of the  Pr$^{3+}$:Y$_2$SiO$_5$ crystal, with 3  hyperfine ground states and 3 excited states. }
\label{AFC}      
\end{figure}
The ultra-narrowband photon pair source described in section
\ref{Sec:SPDCsource:Pr} can be used to generate heralded single
photons compatible with the Praseodymium doped crystal. For
sufficiently high correlations between the two fields, the
detection of an idler telecom photon will indeed project the
signal mode in a single photon Fock state. 

These heralded single photons have been stored in the Pr$^{3+}$:Y$_2$SiO$_5$ as
collective optical atomic excitations using the AFC echo scheme.
The storage device is a 3 mm thick Y$_2$SiO$_5$ sample doped with
a Pr$^{3+}$ concentration of 0.05 $\%$. The relevant optical
transition connects the $^3H_4$ ground state to the $^1D_2$
excited state at a wavelength of 605.977 nm and features a
measured absorption coefficient of 23 cm$^{-1}$, and an
inhomogeneous linewidth of 5 GHz. At zero magnetic field, the
ground state and excited states manifolds are split in 3
metastable states, denoted $\pm 1/2_k$, $\pm
3/2_k$ and $\pm 5/2_k$ (see inset of Fig. \ref{AFC}), where
$k={g,e}$ denotes the ground or the excited state. The spacing
between the hyperfine states is of the order of a few MHz. This
gives an upper limit for the bandwidth as already mentioned. But,
since the separation between ground states is much smaller than
the inhomogeneous broadening, it also creates a complication for
isolating a single class of atoms. A laser sent in the crystal can
indeed be resonant with up to 9 distinct classes of atoms, within
the inhomogeneous broadening. To select a single class of atoms,
the optical pumping scheme first demonstrated by Nilsson et al was
used \cite{Nilsson2004}. After selecting one class of atoms, an
atomic frequency comb is created on the $\pm 1/2_g$ $\rightarrow$
$\pm 3/2_e$ transition  (see Fig. \ref{AFC} for an example of AFC).

One important issue to couple photons to the solid state atomic
ensemble is that the frequency of the photons must be stable
within a few hundred kHz. This can be insured by implementing a
feedback lock system using the laser used to prepare the atomic
frequency comb. One the one hand, the length of the cavity is
locked on the laser with a Pound-Drever-Hall scheme. This insures
that at least one spectral mode emitted by the cavity enhanced
source is resonant with the crystal absorption. On the other hand,
the required double resonance for signal and idler modes is
insured by adjusting the frequency of the pump laser using a
classical signal at the idler wavelength created by difference
frequency generation between the pump laser at 426 nm and the
laser at 606 nm.

In the experiment, a single longitudinal mode was selected for the
idler mode thanks to a Fabry Perot filter cavity. For the heralded
single photon, the crystal itself was used as a filter to prevent
the modes non resonant with the AFC to reach the detector. The probability to have
a single photon in the signal mode before the cryostat conditioned on a detection in
the idler mode (called the heralding efficiency $\eta_H$) was around a few percents in this experiment, limited by dark counts in the idler detector, noise in the idler mode, cavity escape efficiency and optical losses in the signal mode from the cavity to the crystal.

\begin{figure}
\includegraphics[scale=.45]{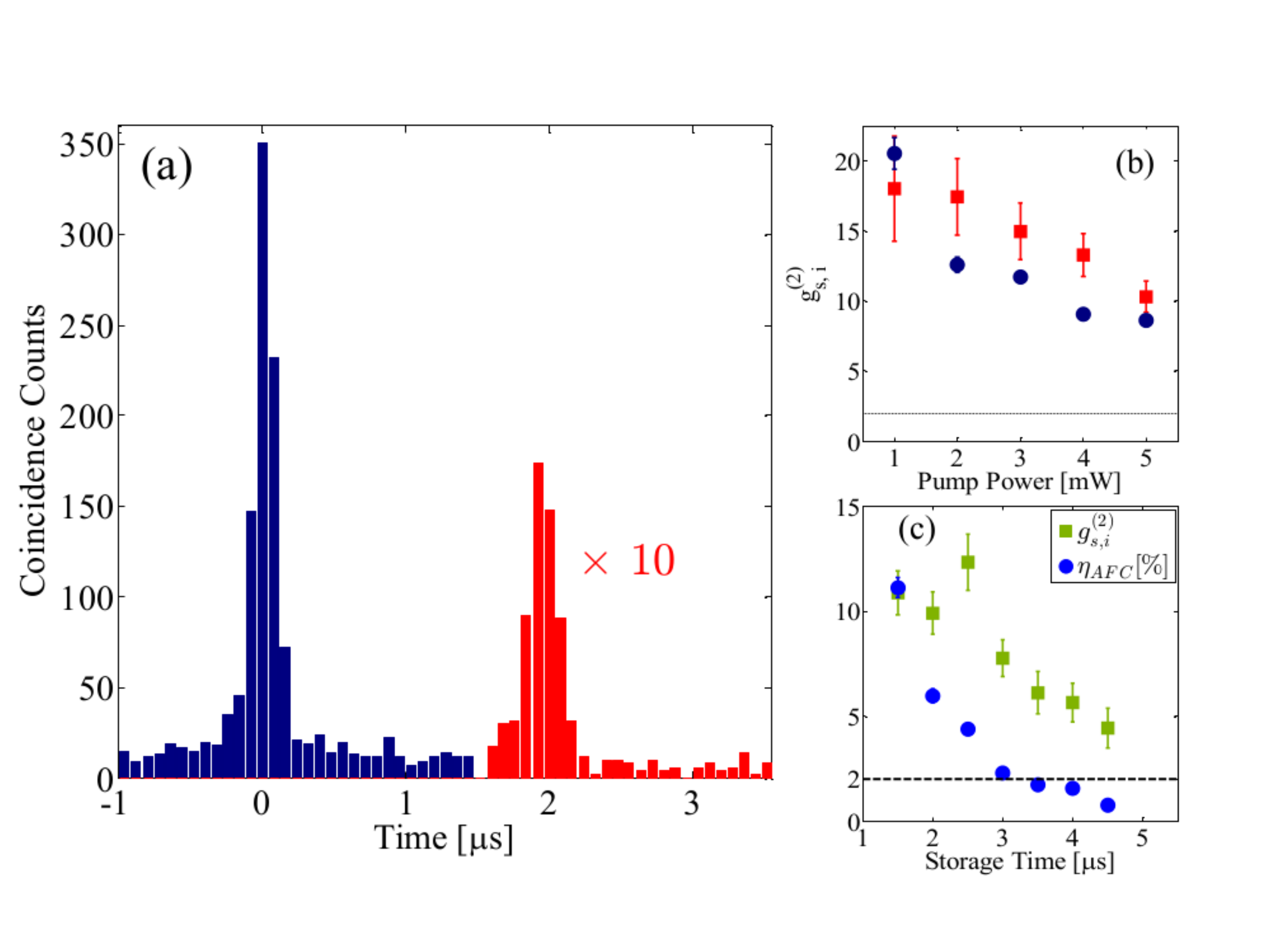}
\caption{ Results of heralded single photon
storage in a Pr$^{3+}$:Y$_2$SiO$_5$ crystal \cite{Rielander2014}. (a) $G^{(2)}_{s,i} (t)$ histogram
without (blue) and with (red) AFC. The preprogrammed storage
time is $2\,\mathrm{\mu s }$ and the power of the
$426.2\,\mathrm{nm}$ pump is $2\,\mathrm{mW}$. (b) The $g^{(2)}_{s,i}$ values as a function of
the pump power for the AFC echo (plain squares) are compared to
those for the input photons (plain circles). 
 The dotted line corresponds to the
classical limit $g^{(2)}_{s,i} = 2$ for two-mode squeezed states. (c) Storage and retrieval efficiency (blue circle) and  $g^{(2)}_{s,i}$  (green squares) as a function of the storage time. For (b) and (c) the data are evaluated for a detection window $\Delta \tau$ = 400 ns and the error bars are evaluated from the raw number of counts assuming Poissonian statistics.}
\label{fig_g2Pr}       
\end{figure}

The heralded single photon at 606 nm was first characterized by
measuring the second order correlation $g^{(2)}_{s,i}(\Delta
\tau)$ function by sending it through the crystal where a 12 MHz
wide transparency window was created. In Fig. \ref{fig_g2Pr}b, the
values of $g^{(2)}_{s,i}(\Delta \tau)$ for the incoming photons
are plotted as a function of the pump power. The value of
$g^{(2)}_{s,i}(\Delta \tau)$ increases when the pump power
decreases, as expected for a two-mode squeezed state. The non
classicality of the input photon was also demonstrated by
violating a Cauchy-Schwarz inequality. The detected coincidence
count rate was around 0.8 Hz per mW of pump power. By correcting
for known optical losses and detection efficiencies, a creation
rate outside the cavity of 2.8 kHz /mW was inferred.

After having confirmed the non classicality of the input light,
the heralded single photon was stored in and retrieved from the
crystal using  atomic frequency comb. In order to avoid the
spurious noise effect described in section \ref{Sec:QStorage:Nd},
the photon pair source pump light was switched off after detection
of an idler photon. The efficiency of the storage and retrieval
process was measured to be up to 10 $\%$ for short storage times.
The second order cross correlation is measured for the stored and
retrieved photon as can be seen in Fig. \ref{fig_g2Pr}a. The values
of $g^{(2)}_{s,i}$ are also plotted as a function of the pump
power in Fig. \ref{fig_g2Pr}b. Surprisingly, the values of $g^{(2)}_{s,i}$ after retrieval
are higher than the one for the input photons, for a big range of
pump powers. This effect has been attributed to the fact that the
memory act as a filter for broadband noise emitted by the photon
pair source. Since the pump light is switched off after the
detection of an idler photon, the atomic frequency comb delays the
signal photon in a noise free region, which therefore increases
the signal to noise ratio and the $g^{(2)}_{s,i}(\Delta \tau)$
\cite{Rielander2014}. This effect highlights that under certain
conditions, quantum memories can act as purifiers by storing only
the signal and not the noise. It has been observed also in other
experiments \cite{McAuslan2012,Maring2014}.

The storage time in the crystal can be chosen by tuning the comb
periodicity $\Delta$. In theory, the minimal $\Delta$ achievable
is given by $2\gamma_hF$, where $\gamma_h$ is the homogeneous
linewidth of the optical transition and $F$ the finesse of the
comb. However, in practice, several effects will limit the
achievable $\Delta$, including power broadening, finite laser
linewidth, spin inhomogeneous broadening, crystal vibrations in
closed loop cryostats, etc. In the present experiment, non
classical correlations between heralding photon and stored and
retrieved heralded photons have been observed until a storage time
of $4.5 \mu s$ (see Fig. \ref{fig_g2Pr}c). This is more than 20 times longer than previous
realizations \cite{Clausen2011,Saglamyurek2011} and would allow
entanglement between crystals separated by km long distance. The
storage and retrieval efficiency dropped by a factor around 10
between 1.5 and 4.5 us (see Fig. \ref{fig_g2Pr}c). This is mainly due to the fact that the
finesse could not be kept constant when decreasing $\Delta$,
because of the minimum achievable width of an absorption peak, due
to the limitations mentioned above. Longer storage times in the
excited state of up to 10 us have recently been obtained with weak
coherent states in Pr$^{3+}$:Y$_2$SiO$_5$ \cite{Maring2014}. Note
also that excited state storage times of up to 30 us have been
obtained using bright pulses storage in a Eu$^{3+}$:Y$_2$SiO$_5$
crystal \cite{Jobez2014}.
\section{Prospects for spin-wave storage with quantum light}
\label{Sec:prospects} In order to increase the storage time and to
achieve an AFC spin wave memory with on demand read-out, it has
been proposed to transfer collective optical atomic excitations to
collective spin excitations (or spin waves), using control fields
(see section \ref{Sec:Protocols}). Proof of principle experiments
have been realized in the classical regime for storage of strong
pulses \cite{Afzelius2010,Gundogan2013,Jobez2014} and we discuss
in this section the prospects to extend this experiments to
quantum light. As mentioned in section \ref{Sec:Protocols}, spin
wave storage requires materials with at least 3 long lived ground
state levels. The best known materials with the required
properties are Pr$^{3+}$ and Eu$^{3+}$ doped crystals.

The main experimental challenge for reaching the quantum regime
 with spin-wave storage is to suppress the noise generated by
the control pulses, which are very close in frequency from the
single photon output (e.g. from 10 to 17 MHz in Pr$^{3+}$:Y$_2$SiO$_5$ and from 35
to 120 MHz in Eu$^{3+}$:Y$_2$SiO$_5$ ). In that respect, one advantage of the AFC
scheme is that the control pulses are temporally separated from
the single photon output, which allow the use of temporal
filtering. However, the control pulses also create transient
phenomena inside the crystal due to e.g. the interaction with
unwanted residual population in the storage state because of
imperfect optical pumping. This can give rise to free induction
decay and fluorescence, which lead to noise emission simultaneous
with the single photon emission. The suppression of this noise
requires narrow spectral filtering, which can be realized using
e.g. a narrow-band optical cavity or a crystal filter
\cite{Zhang2012,Beavan2013}. An interesting figure of merit in the
context of single-photon-level spin-wave storage, taking into
account the noise generated and the storage and retrieval
efficiency,  is the mean number of input photons to achieve a
signal-to-noise ratio (SNR) of 1 in the output mode, denoted as
$\mu_1$. In 2013, an experiment in Eu$^{3+}$:Y$_2$SiO$_5$  achieved $\mu_1$=2.5,
using a Fabry-Perot narrowband filter cavity \cite{Timoney2013}. Very recently, a lower value of  $\mu_1$=0.1 has been  demonstrated in Eu$^{3+}$:Y$_2$SiO$_5$ \cite{Jobez2015}. A low value of  $\mu_1$=0.07 has also been shown in  Pr$^{3+}$:Y$_2$SiO$_5$ \cite{Gundogan2015}, using a reconfigurable transparency window in an another
Pr$^{3+}$:Y$_2$SiO$_5$ crystal as spectral filter.  High fidelity storage of time-bin qubits encoded in weak-coherent states at the single photon level was also demonstrated \cite{Gundogan2015}, allowing the spin-wave memory to operate in the quantum regime.
In order to achieve high
fidelity storage of quantum light, it is important to obtain
$\mu_1\ll 1$. This is because in practice, because of various
optical losses, it is very difficult to obtain a single photon in
front of the memory, with efficiency approaching unity. To achieve
quantum storage with high SNR using the photon pair sources
presented in section \ref{Sec:SPDCsource:General}, the condition
$\mu_1\ll \eta_H$ must be fulfilled. Such an experiment has so far
not been demonstrated.

Beyond the challenge of storing a single photon as a single
spin-wave excitation, another important challenge is to increase
the spin-wave storage time in this single-excitation regime. The
spin-wave memory in a rare-earth-ion doped crystal is limited by
the inhomogeneous spin linewidth \cite{Afzelius2010}. This
limitation can be lifted by utilizing spin-echo techniques to
rephase the inhomogeneous spin dephasing, in which case the
storage time is limited by the spin coherence time. But one can
push the storage time still further by implementing dynamical
decoupling sequences to reach storage times beyond 1 s, a
technique that has been successfully implemented in EIT storage
experiments of strong optical pulses in rare-earth-ion doped
crystals  \cite{Longdell2005,Heinze2013}. However, the use of spin echo tecyhniques for extending the storage time of ensemble-based quantum memories introduces a new potential source of noise. The challenge lies in
avoiding to populate the $|s\rangle$ state with too many atoms,
which would lead to spontaneous emission noise in the output when
reading out the single spin excitation. At first this might appear
to be almost impossible for a single excitation in $|s\rangle$
\cite{Johnsson2004}, but it was shown later that the strong
collective emission into a particular spatial mode of the stored
single excitation provides a very effective spatial filtering of
the spontaneous emission noise \cite{Heshami2011}. The spin-echo
technique must be very efficient, however, to avoid this noise,
and it remains to be seen if the storage fidelity of a single
photon can be high enough when applying spin echo techniques. In a very recent experiment \cite{Jobez2015}  it was shown that spin echo techniques could indeed be used when manipulating an average spin excitation of around 1, in an ensemble composed of 10$^{10}$  Eu$^{3+}$  ions, without adding significant noise to the optical read out of the memory. This was made possible by using a robust and error compensating spin echo sequence which limited the population introduced to the $|s\rangle$ state to well below 1\%. In this way a $\mu_1$ parameter of around 0.3 could be maintained for up to 1 ms of spin-wave storage time, showing that the noise level should in principle allow quantum state storage on a milliseconds time scale. One future goal of this research is to implement dynamical decoupling sequences in order to further extend the spin-wave storage time.

\section{Outlook}Progress to harness the interaction between quantum light and atomic ensembles in a solid state environment has been fast in recent years.  However, the full capabilities of these materials have not yet been exploited in the quantum regime and several challenges remain. For example, an experiment demonstrating simultaneously high efficiency and long storage time of quantum light has not been demonstrated yet.  The quantum networking capabilities also need to be improved. Current research directions include work towards the realization of  long lived  heralded entanglement between remote solid state multimode quantum memories \cite{Simon2007a}. Such an experiment would pave the way to functional elementary segments of quantum repeaters with multiplexed entanglement generation. These applications would strongly benefit from  the realization of a solid state photon pair source with embedded memory \cite{Sekatski2011}.  Another research direction actively pursued is to increase the spectral multiplexing capabilities, together with selective frequency read-out \cite{Sinclair2014,Thiel2014}. Another promising research direction is the integration of these quantum memories with micro and nanostructures, which  would open many interesting opportunities in terms of miniaturization, scalability and integration other optical elements such as quantum light sources and single photon detectors. The coupling of rare-earth ions with nanophotonic structures like  photonic crystal waveguides or cavities would also provide increased light-matter interaction and potentially lead to cavity QED experiments with a low number of rare-earth ions \cite{Zhong2014}. This may also facilitate the detection and manipulation of single rare-earth ions that could be used as quantum bits \cite{Kolesov2012,Utikal2014}. Finally, rare-earth ion doped crystals can be used as spin ensemble that can be coupled to superconductive cavities \cite{Staudt2012,Probst2013}, with the long term goal of connecting superconducting qubits and optical photons.

\begin{acknowledgement}

The results described here has been obtained during several years
of work by the teams at the University of Geneva and ICFO. The
authors would like to particularly acknowledge the contributions
by their co-authors of the work reviewed here, Christoph Clausen,
Imam Usmani, F\'{e}lix Bussi\`{e}res, Nicolas Sangouard, Nicolas
Gisin, Daniel Rielander, Kutlu Kutluer, Mustafa Gundogan, Patrick
Ledingham, Margherita Mazzera, Julia Fekete and Matteo Cristiani.
\end{acknowledgement}



\bibliographystyle{SpringerPhysMWM} 

\printindex
\end{document}